\documentclass{sig-alternate-05-2015}

\usepackage{type1cm}     
\usepackage{graphicx}     
\usepackage{xspace}     
\usepackage{balance}     
\usepackage{booktabs}     
\usepackage{multi row}     
\usepackage[font={bf}, tableposition=top]{caption}     
\usepackage[hyphens]{url}     
\usepackage{bold-extra}     
\usepackage{siunitx}          
\usepackage[vlined,linesnumbered,ruled,noend]{algorithm2e}     
\usepackage[bookmarks, pdftex, colorlinks=false]{hyperref}     
\usepackage[square,numbers]{natbib}     
\usepackage{microtype}    
\usepackage{units}     
\usepackage{mathtools}     
\usepackage{amssymb}     
\usepackage[show]{chato-notes}
\usepackage{relsize}
\usepackage{graphicx}
\usepackage{caption}
\usepackage{booktabs}     
\usepackage{multirow}
\usepackage{mathtools} 
\usepackage[]{inputenc}
\usepackage{amssymb}
\usepackage{amsmath}
\usepackage{siunitx} 
\usepackage[]{algorithm2e}
\usepackage[font={small}]{subfig} 

\makeatletter 
\def\@copyrightspace{\relax}
\makeatother

\newtheorem{theorem}{Theorem}
\newtheorem{lemma}{Lemma}

\newtheorem{problem}{Problem}

\newcommand{\spara}[1]{\smallskip\noindent\textbf{#1}}
\newcommand{\mpara}[1]{\medskip\noindent\textbf{#1}}

\newcommand{\kedgeadd}{{\ensuremath{k}}-{\sc Edge\-Addition}}
\newcommand{\kedgeaddexp}{{\ensuremath{k}}-{\sc Edge\-Addition\-Expectation}}

\newcommand{\rwc}{\ensuremath{\text{\sc RWC}}}
\newcommand{\rov}{\ensuremath{\text{\sc ROV}}}
\newcommand{\rovap}{\ensuremath{\text{\sc ROV-AP}}}

\newenvironment {squishlist}
{\begin{list}{$\bullet$}
  { \setlength{\itemsep}{0pt}
     \setlength{\parsep}{3pt}
     \setlength{\topsep}{3pt}
     \setlength{\partopsep}{0pt}
     \setlength{\leftmargin}{1.5em}
     \setlength{\labelwidth}{1em}
     \setlength{\labelsep}{0.5em} } }
{\end{list}}

\begin{document}

\title{Reducing Controversy by Connecting Opposing Views\thanks{Accepted at WSDM 2017. Please cite the WSDM 2017 version of this paper.}}
\author{
Kiran Garimella\\
       \affaddr{Aalto University}\\
       \affaddr{Helsinki, Finland}\\
       \email{kiran.garimella@aalto.fi}
\and
Gianmarco De~Francisci~Morales\\
      \affaddr{Qatar Computing Research Institute}\\
      \affaddr{Doha, Qatar}\\
      \email{gdfm@acm.org}
\and  
Aristides Gionis\\
      \affaddr{Aalto University}\\
      \affaddr{Helsinki, Finland}\\
      \email{aristides.gionis@aalto.fi}
\and
Michael Mathioudakis\\
       \affaddr{HIIT, Aalto University}\\
       \affaddr{Helsinki, Finland}\\
       \email{michael.mathioudakis@hiit.fi}
}

\maketitle

\begin{abstract}
Society is often polarized by controversial issues, that split the population into groups of opposing views.
When such issues emerge on social media, we often observe the creation of `echo chambers', i.e., situations where like-minded people reinforce each other's opinion, but do not get exposed to the views of the opposing side.
In this paper we study algorithmic techniques for bridging these chambers, and thus, reducing controversy. 
Specifically, we represent the discussion on a controversial issue with an \emph{endorsement graph}, and cast our problem as an {\em edge-recommendation problem} on this graph.
The goal of the recommendation is to reduce the \emph{controversy score} of the graph, which is measured by a recently-developed metric based on random walks.
At the same time, we take into account the \emph{acceptance probability} of the recommended edge, which represents how likely the edge is to materialize in the endorsement graph.

We propose a simple model based on a recently-developed user-level controversy score, that is competitive with state-of-the-art link-prediction algorithms.
We thus aim at finding the edges that produce the largest reduction in the controversy score, in expectation.
To solve this problem, we propose an efficient algorithm, which considers only a fraction of all the combinations of possible edges.
Experimental results show that our algorithm is more efficient than a simple greedy heuristic, while producing comparable score reduction.
Finally, a comparison with other state-of-the-art edge-addition algorithms shows that this problem is fundamentally different from what has been studied in the literature.
\end{abstract}


\section{Introduction}

Polarization around controversial issues is a well studied phenomenon in the social sciences~\citep{isenberg1986group,sunstein2002law}.
Social media have arguably facilitated the emergence of such issues, with the scale of discussions and the publicity they foster. 
In this paper, we study how to reduce the polarization in controversial issues on social media by creating bridges across opposing sides.

We focus on controversial issues that create discussions online.
Usually, these discussions involve a fair share of ``retweeting'' or ``sharing'' opinions of authoritative figures that the user agrees with.
Therefore, it is natural to model the discussion as an \emph{endorsement graph}:
a vertex $v$ represents a user, and a directed edge $(u,v)$ represents the fact that user $u$ endorses the opinion of user $v$.

Given this modus operandi, and the existence of confirmation bias, homophily, selective exposure, and related social phenomena in human activities, the existence of echo chambers online is not surprising~\citep{garrett2009echo, del2015echo}.
The existence of these chambers is a hindrance to the democratic process and to the functioning of society at large, as they cultivate isolation and misunderstanding between opposing sides.

A solution to this problem is to create \emph{bridges} that connect people of opposing views.
By putting different parts of the endorsement graph in contact, we hope to reduce the polarization of the graph.

We operationalize this concept by leveraging recent advances in quantifying online controversy~\citep{garimella2016quantifying}.
In particular, given a metric that measures how controversial an issue discussed on social media is, our goal is to find a small number of edges, called bridges, that minimize this measure.
That is, we seek to propose (content produced by) a user $v$ to another user $u$, hoping that $u$ endorses $v$ by spreading her opinion.
This action would create a new edge (a bridge) in the endorsement graph, thus reducing the controversy score of the graph itself.

Clearly, some bridges are more likely than others to materialize than others. 
For instance, people in the `middle' might be easier to convince than people on the two extreme ends of the political spectrum~\cite{liao2014can}.
We take this issue into account by modeling an \emph{acceptance probability} for a bridge as a separate component of the model.
This component can be implemented by any generic link-prediction algorithm that gives a probability of materialization to each non-existing edge.
In addition, we propose a simple model based on a recently developed user-level controversy score~\citep{garimella2016tscpaper} which nicely captures the dynamics and properties of the endorsement graph.
Therefore, we seek to find bridges that minimize the \emph{expected} controversy score.

The core of this paper is an algorithm to solve the aforementioned problem.
We show that a brute-force approach is not only unfeasible, as it requires one to evaluate a combinatorial number of candidates, but also unnecessary.
Moreover, our algorithm needs to consider far fewer than the $\mathcal{O}(n^2)$ possible edges (where $n$ is the number of vertices in the graph) needed by a simple greedy heuristic.

\begin{figure*}[t]
\begin{minipage}{.24\linewidth}
\centering
\subfloat[]{\label{}\includegraphics[width=\textwidth, height=\textwidth]{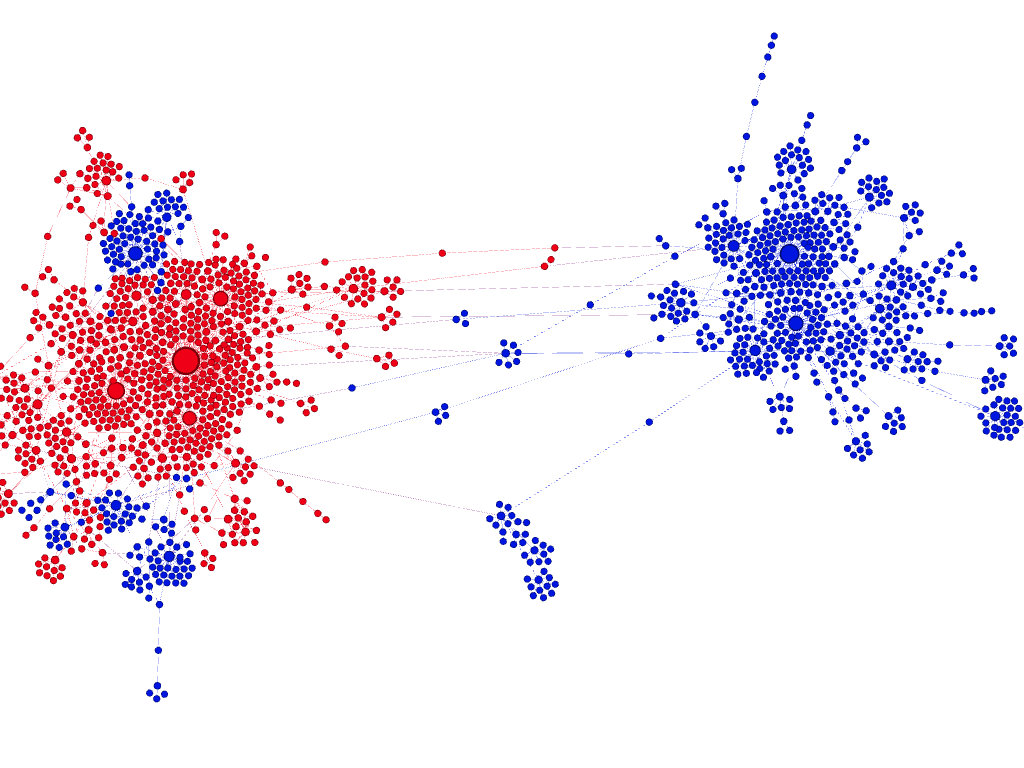}}
\end{minipage}%
\begin{minipage}{.24\linewidth}
\centering
\subfloat[]{\label{}\includegraphics[width=\textwidth, height=\textwidth]{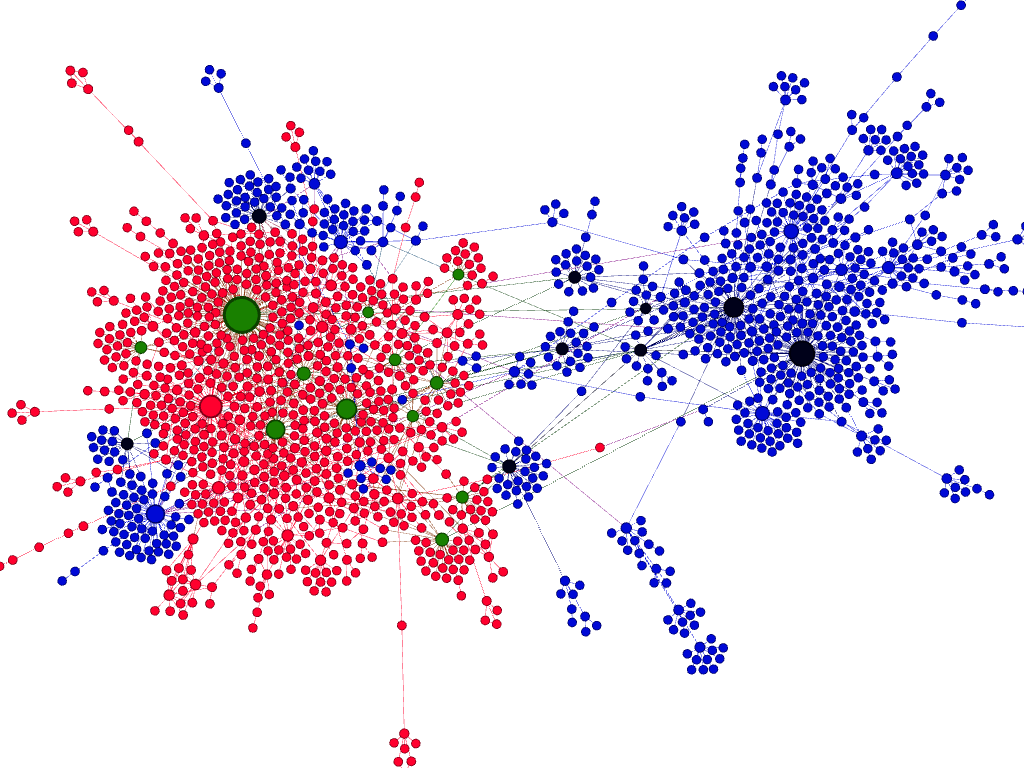}}
\end{minipage}
\begin{minipage}{.24\linewidth}
\centering
\subfloat[]{\label{}\includegraphics[width=\textwidth, height=\textwidth]{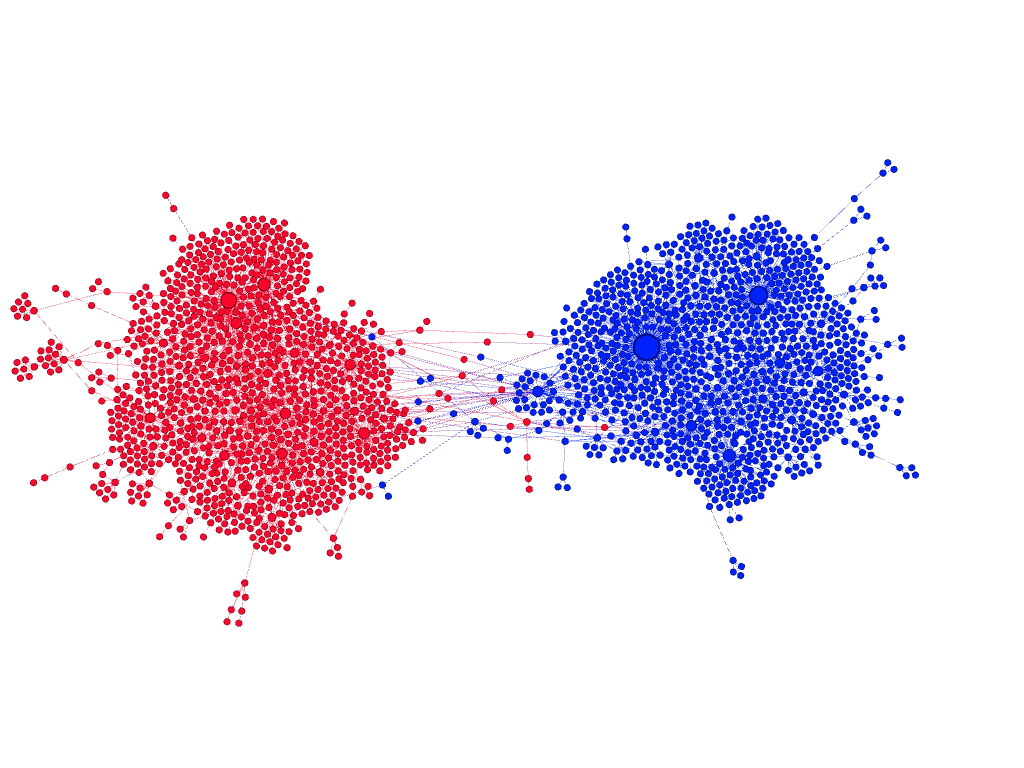}}
\end{minipage}
\begin{minipage}{.24\linewidth}
\centering
\subfloat[]{\label{}\includegraphics[width=\textwidth, height=\textwidth]{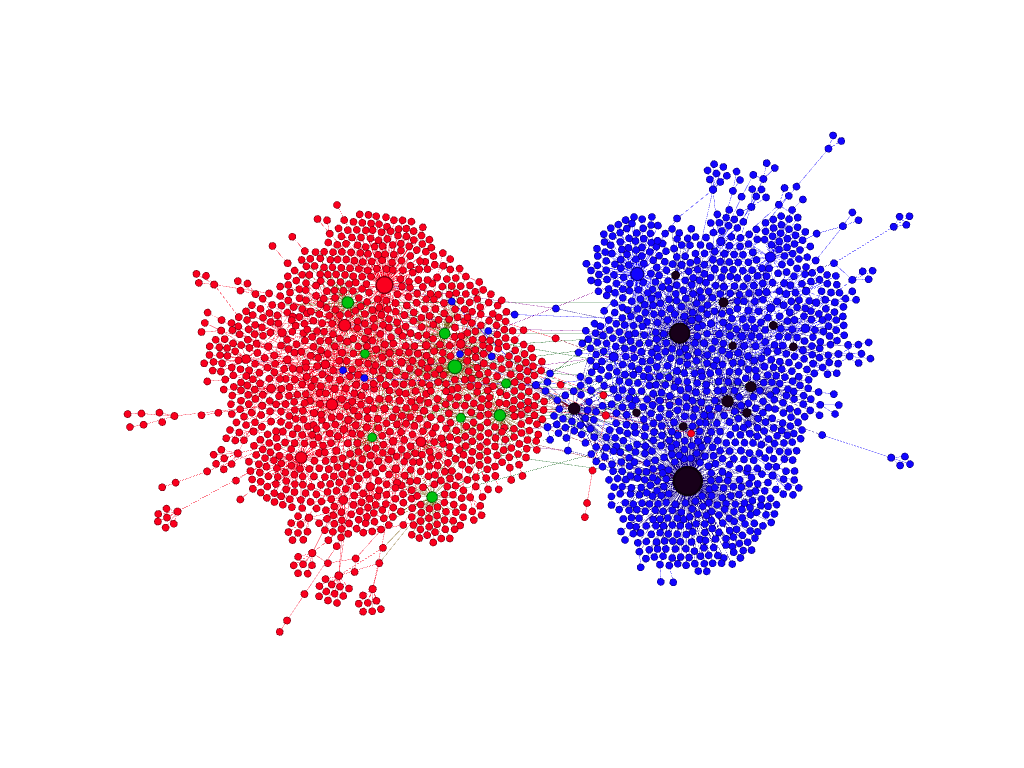}}
\end{minipage}
\par\medskip
\caption{Sample retweet graphs (visualized using the force-directed layout algorithm in Gephi) for \#beefban (a,b) and \#russia\_march (c,d) before (a,c) and after (b,d) the addition of 20 edges using \rovap. We clearly see that the two clusters appear to come closer after the edge addition.}
\label{fig:visualization}
\vspace{-\baselineskip}
\end{figure*}

Experimental results show that our algorithm is able to minimize the controversy score of a network efficiently and as effectively as the greedy algorithm. An example of the results produced by our algorithm are shown in Figure~\ref{fig:visualization}. We see that the two sides of the controversy appear to come closer upon adding the edges proposed by our algorithm. For more details, refer to Section~\ref{sec:casestudy}.

In addition, our experiments show that previously-proposed methods for edge addition that optimize for different objective functions are not applicable to the problem at hand.

In summary, our contributions are the following:
\begin{squishlist}
\item We study the problem of bridging echo chambers algorithmically, in a language and domain agnostic way for the first time;
Studies in the past that tried to address this problem focus mostly on understanding {\em how} 
to recommend content to an ideologically opposite side, while our focus is on {\em who} to recommend opposing-view content to.
We believe that the two approaches complement each other 
in bringing us closer to bursting the filter bubble.
\item We build on top of results from recent user studies~\cite{munson2013encouraging,liao2014expert,vydiswaran2015overcoming} on how users prefer to consume content from opposing views and formulate the task as 
an edge-recommendation problem in an endorsement graph, 
that also takes into account the acceptance probability of a recommendation;
\item We provide a method to estimate the acceptance probability of a recommendation that fits well in this setting;
\item We propose an efficient algorithm to solve the problem, which considers fewer candidates than a greedy baseline;
\item We extensively evaluate the proposed algorithm on real-world data, 
and demonstrate that it outperforms all sensible baselines.
\end{squishlist}



\section{Related Work}
\label{section:related-work}

\spara{Making recommendations to decrease polarization.}
The web offers the opportunity 
to easily access any kind of information. 
Nevertheless, several studies have observed that, 
when offered choice, 
users prefer to be exposed to agreeable and like-minded content. 
For instance, Liao et al.~\cite{liao2013beyond} report that 
``{\em even when opposing views were presented side-to-side, 
people would still preferentially select information that reinforced their existing attitudes.}''
This selective-exposure phenomenon 
(also called ``filter bubble'' or ``echo chamber'') 
has led to increased fragmentation and polarization online. 
A wide body of recent studies have 
studied~\cite{adamic2005political,conover2011political,mejova2014controversy} and 
quantified~\cite{akoglu2014quantifying,garimella2016quantifying,guerra2013measure,morales2015measuring} 
this divide.

Given the ill-fated consequences of polarization on the society~\cite{pariser2011filter,sunstein2009republic}, 
it is well-worth investigating whether online polarization and filter bubbles can be avoided.
One simple way to achieve this is to 
``nudge'' individuals towards being exposed to opposing view-points,
an idea that has motivated several pieces of work in the literature.

Liao et al.~\cite{liao2014can,liao2014expert} 
attempt to limit the {\em echo-chamber effect} 
by making users aware of other users' stance on a given issue, 
the extremity of their position, and their expertise. 
Their results show that participants who seek to acquire more accurate information about an issue 
are exposed to a wider range of views, and 
agree more with users who express moderately-mixed positions on the issue. 

Vydiswaran et al.~\cite{vydiswaran2015overcoming} perform a user study aimed 
to understand ways to best present information about controversial issues to users so as to persuade them. 
Their main relevant findings reveal that factors 
such as showing the credibility of a source, or the expertise of a user, 
increases the chances of other users believing 
in the content.
In a similar spirit, \cite{munson2013encouraging} create a browser widget that measures 
and displays the bias of users based on the news articles they read.
Their study concludes that displaying to users their bias helps them read articles of opposing views.

Graells-Garrido et al.~\cite{graells2013data} 
show that mere display of opposing-view content has negative emotional effect.
To overcome this effect, they 
propose a visual interface for making recommendations from a diverse pool of users, 
where diversity is with respect to user stances on a topic.
In contrast,
Munson et al.~\cite{munson2010presenting} 
show that not all users value diversity and that the way of presenting information 
(e.g., highlighting vs.\ ranking) makes a difference in the way users perceive information.
In a different direction, 
Graells-Garrido et al.~\cite{graells2014people} propose to find ``intermediary topics'' 
(i.e., topics that may be of interest to both sides) 
by constructing a {\em topic graph}.  
They define intermediary topics to be those topics 
that have high betweenness centrality and topic diversity.


\smallskip
Based on the papers discussed above, we make the following observations:

\smallskip
\noindent
(a) Though a lot of studies have been proposed to solve the problem of decreasing polarization, 
there is a lack of an algorithmic approach
that works in a domain- and language-independent manner.
Instead, the approaches listed above are mostly based on user studies and/or hand-crafted datasets.
To our knowledge, 
this paper is the first to offer such an algorithmic approach.

\smallskip
\noindent
(b) Additionally, the studies discussed above focus mostly on understanding {\em how} 
to recommend content to an ideologically opposite side. 
Instead, the approach presented in this paper deals with the problem
of finding {\em who} to recommend opposing-view content.
We believe that combining the two approaches 
can bring us a step closer to bursting the filter bubble.

\smallskip
\noindent
(c) The studies discussed above suggest that 
($i$) it is possible to nudge people by recommending content from an opposing side~\cite{munson2013encouraging}, 
($ii$) extreme recommendations might not work~\cite{graells2014people},
($iii$) people ``in the middle'' are easier to convince~\cite{liao2014can},
($iv$) expert users and hubs are often less biased and 
can play a role in convincing others~\cite{liao2014expert,vydiswaran2015overcoming}

\smallskip
In the design of our algorithm
we explicitly take into account the considerations ($i$)--($iv$).

\spara{Adding edges to modify the graph structure.}
In addition to the work on explicitly reducing polarization in social media, 
there are many papers aiming to make a network more cohesive by edge additions, 
where cohesiveness is quantified using graph-theoretic properties, 
such as 
shortest paths~\cite{parotisidis2015selecting,papagelis2011suggesting},
closeness centrality~\cite{parotsidis2016centrality},
diameter~\cite{demaine2010minimizing}, 
eccentricity~\cite{perumal2013minimizing},
communicability~\cite{arrigo2015edge,arrigo2016updating},
synchronizability~\cite{zeng2012manipulating}, 
 and 
natural connectivity~\cite{chan2014make}.

The paper that is is conceptually closest to ours is the one by Tong et al.~\cite{tong2012gelling}, 
which aims to add and remove edges in a graph to reduce the dissemination of content (e.g., viruses).
The proposed approach is to try to maximize the largest eigenvalue, 
which determines the epidemic threshold and, thus,  
the properties of information dissemination in networks.

\smallskip
The similarity of the above-mentioned approaches to our paper 
is limited to the fact that the goal is to modify a graph by edge additions. 
However, the proposed approach and objective function is 
predominantly different than those found in other methods. 



\section{Preliminaries and \\ problem definition}
\label{sec:rwc}

To ensure an algorithmic approach to identifying controversial issues
and selecting which edges to recommend in order to reduce controversy in the social network, 
we need to rely on a measure of controversy.
As reviewed in Section~\ref{section:related-work}
there are several measures of quantifying controversy in social media~\cite{adamic2005political,akoglu2014quantifying,conover2011political,garimella2016quantifying,mejova2014controversy,morales2015measuring}.
%
In this paper, we adopt the controversy measure proposed by Garimella et al.~\cite{garimella2016quantifying},
as it is the most recent work
and it was shown to work reliably in multiple domains; 
in contrast, other measures focus on a single topic (usually politics) 
and/or require domain-specific knowledge.
The adopted controvery measure 
consists of the following steps~\cite{garimella2016quantifying}:

\smallskip
\noindent
($i$)
Given a topic $t$ for which we want to quantify its controversy level, 
we create an {\em endorsement graph} $G=(V,E)$.
This is a graph between users who have generated content relevant to~$t$.
For instance, if $t$ is specified by a hashtag, 
the nodes of the endorsement graph is the set of all users who
have used this hashtag. 
The edges of the endorsement graph are defined to be 
{\em retweets} among the users, 
in order to capture user-to-user endorsement. 

\smallskip
\noindent
($ii$)
The nodes of the endorsement graph $G=(V,E)$ are partitioned into two disjoint sets $X$ and $Y$, 
i.e., $X \cup Y = V$ and $X \cap Y = \emptyset$.
The partitioning is based on the graph structure
and it is obtained using any graph-partitioning algorithm. 
The intuition is that for controversial topics (as shown in~\cite{garimella2016quantifying}),
the partitions $X$ and $Y$ are well separated
and correspond to the opposing sides of the controversy.

\smallskip
\noindent
($iii$)
The last step of computing the controversy measure 
relies on a {\em random walk}.
In particular, the measure, which is named {\rwc} (random-walk controversy) score, 
is defined as the difference of the probability that a random walk 
starting on one side of the partition will stay on the same side 
and the probability that the random walk will cross to the other side.
This is computed using two personalized PageRank computations, 
where the probability of restart is set at a random node on each side,
and the final probability is taken by considering the stationary distribution of only the high-degree nodes.

In more detail, 
let $P$ be the column-stochastic transition probability matrix for the random walk, 
and let $X^*$ and $Y^*$ be a set of the $k_1$, $k_2$ highest in-degree nodes of the two partitions $X$ and $Y$ respectively. 
Let $r_{_X}$ be the personalized PageRank vector 
for the random walk starting in $X$ with restart vector $e_{_X} = \mathrm{uniform}(X)$
and restart probability $(1 - \alpha) \in [0, 1]$.
Similarly, let $r_{_Y}$ be the personalized PageRank vector for the random walk starting in $Y$ 
with restart vector $e_{_Y} = \mathrm{uniform}(Y)$ and restart probability $(1 - \alpha)$.

Let $P_{_X}$ and $P_{_Y}$ be the transition matrices corresponding 
to the two random walks starting from the corresponding side. 
Note that if there are no dangling nodes in the graph then $P_{_X} = P_{_Y} = P$. 
In the case of dangling nodes, following standard practice,
the matrices $P_{_X}$ and $P_{_Y}$ are defined
so that the transition probabilities from the dangling nodes 
are equal to the restart vectors $e_{_X}$ and $e_{_Y}$, respectively.
The personalized PageRank for the two random walks (starting in $X$ and starting in $Y$)
is given by equations:
\begin{equation}
 \left.\begin{aligned}
    r_{_X} = \alpha P_{_X} r_{_X} + (1 - \alpha) e_{_X} \\
    r_{_Y} = \alpha P_{_Y} r_{_Y} + (1 - \alpha) e_{_Y}.
     \end{aligned}
 \right.
 \label{eq:rxry}
\end{equation}
Let $c_{_X}$ be a vector of size $n$ having value 1 in the coordinates
that correspond to the high-degree nodes $X^*$ and 0 elsewhere, 
and similarly define $c_{_Y}$.
The random-walk controversy score ${\rwc}(G,X,Y)$ is defined as:
\begin{eqnarray}
\label{eq:rwc-simplified1}
\begin{split}
 {\rwc}(G,X,Y) & = (\transpose{c_{_X}} r_{_X}  + \transpose{c_{_Y}} r_{_Y}) - (\transpose{c_{_Y}} r_{_X}  + \transpose{c_{_X}} r_{_Y} ) & \\
 & = \transpose{(c_{_X} - c_{_Y})} (r_{_X} - r_{_Y}).  &
\end{split}
\end{eqnarray}
Using Equations~(\ref{eq:rxry}), the Equation~(\ref{eq:rwc-simplified1}) can be written as:

\begin{equation*}
\begin{split}
{\rwc}&(G,X,Y) = \\
& (1-\alpha) \transpose{(c_{_X} - c_{_Y})} ((I - \alpha P_{_X})^{-1}e_{_X} - (I - \alpha P_{_Y})^{-1}e_{_Y}),
\end{split}
\end{equation*}
or
\begin{equation}
\label{eq:rwc_simplified2}
{\rwc}(G,X,Y) = (1-\alpha) \transpose{(c_{_X} - c_{_Y})} (M_{_X}^{-1}e_{_X} - M_{_Y}^{-1}e_{_Y}),
\end{equation}
for $M_{_X} = (I - \alpha P_{_X})$ and $M_{_Y} = (I - \alpha P_{_Y})$.

Given the controversy measure $\rwc(G)$, 
the problem we consider in this paper can be formulated as follows. 

\begin{problem}[\kedgeadd]
Consider a graph $G(V,E)$
whose nodes are partitioned into two disjoint sets $X$ and $Y$ 
{\em (}$X \cup Y = V$ and $X \cap Y = \emptyset${\em )}, 
and an integer $k$.
Find a set of $k$ edges $E' \subseteq V \times V \setminus E$ 
to add to $G$ and obtain a new graph $G' = (V,E\cup E')$, 
so that the controversy score $\rwc(G',X,Y)$ is minimized.
\label{problem:basic}
\end{problem}

Note that the two partitions $X$ and $Y$ are considered fixed and part of the input. 
Also the high-degree nodes on which the score depends are considered fixed between $G$ and $G'$.



\section{Algorithms}
\label{sec:algorithms}

A brute-force approach to solve the problem would be to consider all 
$\mathcal{O}(\binom{n^2}{k})$ combinations of $k$ possible edges to add.
A greedy approach would be to select $k$ edges in $k$ steps, 
each time evaluating the improvement in the value of \rwc\ among the remaining $\mathcal{O}(n^2)$ edges.
Even for the greedy approach, though, 
the number of possible edges to consider is prohibitively high in real settings.
Since computation of the polarity score is an expensive operation, 
we would like to invoke the polarity-score function as few times as possible.
That is, we aim to consider far fewer candidate edges --- 
ideally sub-linear in real-world settings.

At a high level, the algorithm we propose works as follows.
It considers only the edges between the high-degree nodes of each side.
For each such edge, it computes the reduction in the \rwc\ score obtained when that edge is added to the original network.
It then selects the $k$ edges that lead to the lowest score when added to the network individually.

\hrulefill

\subsubsection*{Exemplary case}

To motivate the proposed algorithm, we study an exemplary case.
We use this case to justify why we opt to add edges to connect high-degree nodes across the two sides.

\begin{figure}
\begin{center}
\includegraphics[width=0.35\textwidth]{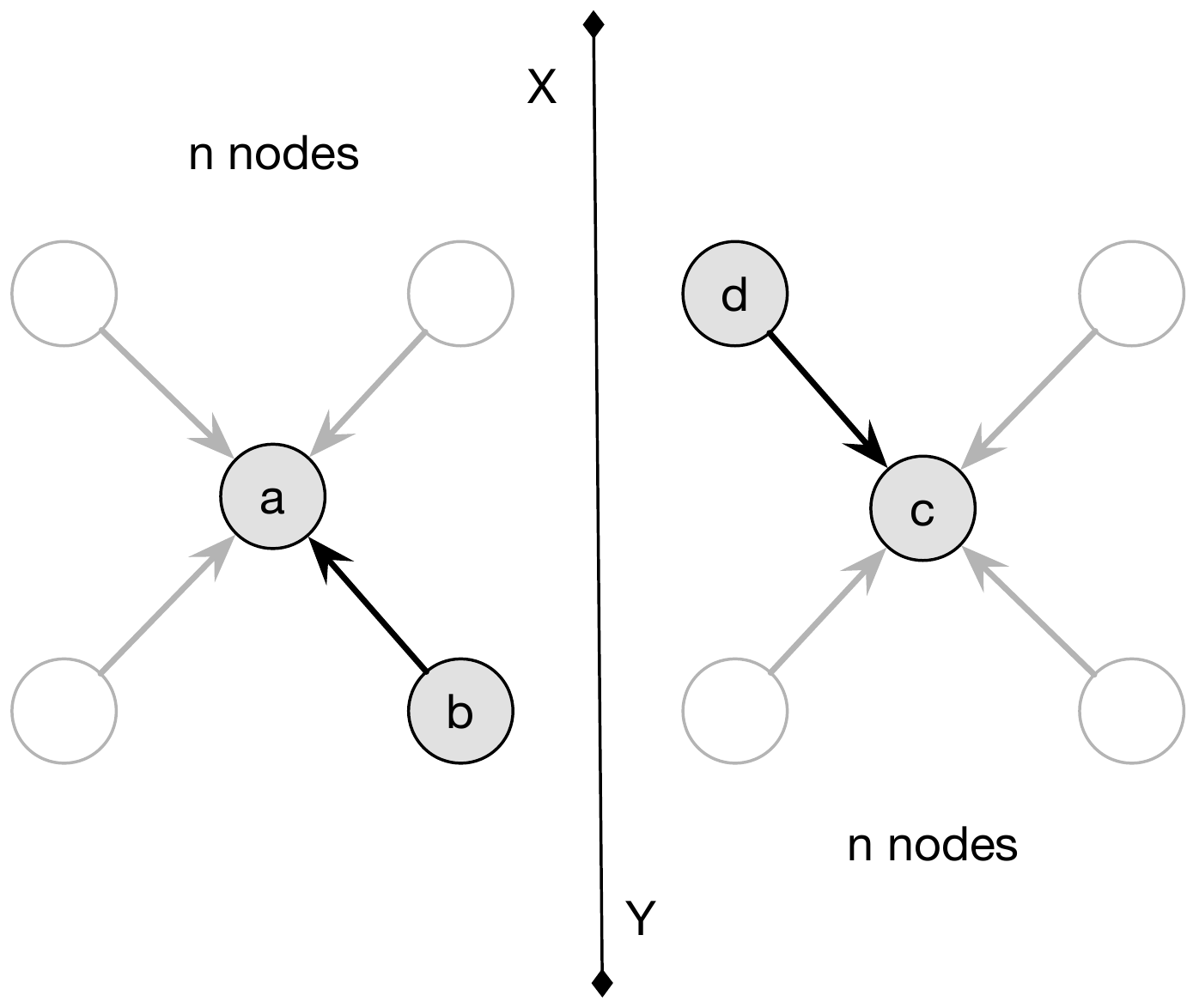}
\caption{Exemplary case, of a network that consists of two disjoint star-like graphs, each of size $n$. We wish to add one directed edge so as to minimize the resulting \rwc\ score.}
\label{fig:two_stars}
\end{center}
\vspace{-\baselineskip}
\end{figure}

Consider a hypothetical network represented by the directed graph shown in Figure~\ref{fig:two_stars}.
The graph consists of two disjoint stars, each comprised of $n$ nodes.
Intuitively, each star represents one side of the controversy.
The center of each star is the high degree nodes of each side.
Following the definition of Problem~\ref{problem:basic} for $k = 1$, we ask which directed edge we should add in order to minimize the controversy score \rwc\ of the entire network.

Without loss of generality, we consider the following four cases of edges: (i) from $a$ to $c$, (ii) from $a$ to $d$, (iii) from $b$ to $c$, (iv) from $b$ to $d$.
Among the four edges, it is the first one, from $a$ to $c$, that connects the two centers of the two stars.
We can analytically formulate the \rwc\ score we obtain when each of these edges is added to the original network, denoted with $s_{a\rightarrow c}$, $s_{a\rightarrow d}$, $s_{b\rightarrow c}$, $s_{b\rightarrow d}$, respectively.
The respective \rwc\ scores are given by the following formulas (details omitted due to lack of space):
\begin{equation*}
	s_{a\rightarrow c} =
	\frac{(-\alpha^2 + \alpha)\cdot n  + (\alpha - 1)^2}{(\alpha^2 + \alpha + 1)\cdot n - \alpha^2} + \frac{\alpha\cdot n - \alpha + 1}{(\alpha + 1)\cdot n - \alpha}
\end{equation*}
\begin{equation*}
	s_{a\rightarrow d} = 
	\frac{(-\alpha^3 + \alpha)\cdot n + \alpha^3 - \alpha^2 - \alpha + 1}{(\alpha^3 + \alpha^2 + \alpha + 1)\cdot n - \alpha^3} + \frac{\alpha\cdot n - \alpha + 1}{(\alpha + 1)\cdot n - \alpha}
\end{equation*}
\begin{equation*}
	s_{b\rightarrow c} =
	\frac{2\alpha\cdot n - 3\alpha + 2}{(\alpha + 1)\cdot n - \alpha}
\end{equation*}
\begin{equation*}
	s_{b\rightarrow d} =
	\frac{\alpha\cdot n - \alpha + 1}{(\alpha + 1)\cdot n - \alpha} + \frac{2\alpha\cdot n - 3\alpha - \alpha^2 + 2}{2(\alpha + 1)\cdot n + \alpha^2 - 2\alpha}
\end{equation*}

\begin{theorem}
For $n \rightarrow \infty$, $\alpha\in [0,1]$, we have
$$s_{a\rightarrow c} \leq s_{a\rightarrow d}, s_{b\rightarrow c}, s_{b\rightarrow d}.$$
\label{theorem:best_edge}
\end{theorem}
\proof{
We have
\begin{equation*}
	s_{a\rightarrow c} \underset{n\rightarrow\infty}{\rightarrow} \frac{-\alpha^2 + \alpha}{\alpha^2 + \alpha + 1} + \frac{\alpha}{\alpha + 1}
\end{equation*}
\begin{equation*}
	s_{a\rightarrow d} \underset{n\rightarrow\infty}{\rightarrow} \frac{-\alpha^3 + \alpha}{\alpha^3 + \alpha^2 + \alpha + 1} + \frac{\alpha}{\alpha + 1}
\end{equation*}
\begin{equation*}
	s_{b\rightarrow c} \underset{n\rightarrow\infty}{\rightarrow} \frac{2\alpha}{\alpha + 1}
\end{equation*}
\begin{equation*}
	s_{b\rightarrow d} \underset{n\rightarrow\infty}{\rightarrow} \frac{\alpha}{\alpha + 1} + \frac{2\alpha}{2(\alpha + 1)} = \frac{2\alpha}{\alpha + 1}
\end{equation*}
and the inequalitites follow trivially.\qed}

Therefore, the edge from node $a$ to node $c$ is the one that leads to minimum score.
Theorem~\ref{theorem:best_edge} provides the optimal edge for a special case.
Even though real networks do not match this case exactly, they often have a structure that resembles star-graphs in certain ways: a small number of highly popular nodes receive incoming edges (retweets) from a large number of other nodes.
It is based on the following model: In a controversial setting, there are thought leaders and followers. 
Most activity in the endorsement graph happens around retweeting and spreading the voice of the leaders across their side, on each side. This leads to a polarized structure which looks like a union of stars on each side of the controversy. 
The theorem suggests intuitively that edges between high-degree nodes of either side are good candidates to add to obtain a low \rwc\ score.

\hrulefill

The exemplary case described above motivates us to consider edges between high-degree nodes from either side.
The algorithm for selecting the edges to be added is shown as Algorithm~\ref{alg:algorithm1}.
Its running time is $\mathcal{O}(k_1\cdot k_2)$, where $k_1$, $k_2$ are the number of high-degree nodes chosen in X and Y respectively.

\begin{algorithm}
\LinesNumbered
 \KwIn{Graph G, number of edges to add, $k$; $k_1,k_2$ high degree nodes in $X,Y$ respectively}
 \KwOut{List of $k$ edges that minimize the objective function, \rwc}
 Initialize: Out $\leftarrow$ {\it empty list} \;
 \For{i = 1:$k_1$}{
  node $u$ = X[i]\;
  \For{j = 1:$k_2$}{
   node $v$ = Y[j]\;
   Compute $\delta \rwc_{u\rightarrow v}$, the decrease in \rwc\ if the edge (u, v) is added\;
   Append $\delta \rwc_{u\rightarrow v}$ to Out\;
   Compute $\delta \rwc_{v\rightarrow u}$, the decrease in \rwc\ if the edge (v, u) is added\;
   Append $\delta \rwc_{v\rightarrow u}$ to Out\;
   }
 }
 sorted $\leftarrow$ sort(Out) by $\delta \rwc$ by decreasing order \;
 \Return top k from sorted;
 \caption{Algorithm for \kedgeadd}
 \label{alg:algorithm1}
\end{algorithm}

\subsection{Incorporating Acceptance Probabilities}
\label{sec:acceptance}

Problem~\ref{problem:basic} is formulated to seek the edges that lead to the lowest \rwc\ score {\it if added} to the network.
In a recommendation setting, however, the selected edges do not always materialize 
(e.g., the recommendation might be rejected by the user).
In such settings, it makes sense to choose the edges that minimize the \rwc\ score {\it in expectation}, under a probabilistic model $\mathbb{A}$ that provides the probability that a set of edges are accepted once recommended. This leads us to Problem~\ref{problem:expectation}.

\begin{problem}[\kedgeaddexp]
Consider a graph $G=(V,E)$
whose nodes are partitioned into two disjoint sets $X$ and $Y$ 
{\em (}$X \cup Y = V$ and $X \cap Y = \emptyset${\em )}, 
and an integer~$k$.
Find a set of $k$ edges $E' \subseteq V \times V \setminus E$ 
to add to $G$ and obtain a new graph $G' = (V,E\cup E')$, 
so that the expected controversy score $E_{_A}[\rwc(G',X,Y)]$ is minimized
under acceptance model $\mathbb{A}$.
\label{problem:expectation}
\end{problem}


We build such an acceptance model $\mathbb{A}$ on the feature of {\it user polarity} from~\cite{garimella2016tscpaper}.
Intuitively, this polarity of a user that takes values in the interval $[-1, 1]$, captures how much the user belongs to either side of the controversy.
High absolute values (close to $-1$ or $1$) indicate that the user clearly belongs to one side of the controversy, while middle values (close to $0$) indicate that the user is in the middle of the two sides.
The reason we employ user-polarity as a feature is that, intuitively, we expect that users of different sides have endorsed content from different sides with varying probability, 
and this probability is a good proxy for how likely it is for an edge to exist. 
For example, a user with polarity close to -1 is more likely to endorse a user with a negative polarity than, 
say, a user with a polarity +1. 

Technically, the polarity score $R_u$ of user $u$ is defined as follows.
Let $l_u^{_X}$ and $l_u^{_Y}$  be the expected time a random walk needs to hit the high degree nodes of side $X$ and $Y$, respectively, starting from node $u$.
Moreover, let $\rho^{X}(u) \in [0,1]$ and $\rho^{Y}(u) \in [0,1]$ be the fraction of other nodes $u'$ for which $l_{u'}^{_X} < l_{u}^{_X}$ and $l_{u'}^{_Y} < l_{u}^{_Y}$, respectively.
The polarity of user $u$ is then defined as 
\begin{equation}
	R_u = \rho^{X}(u) - \rho^{Y}(u) \;\;\in [-1, 1].
\end{equation}

Now let $u$ and $v$ be two users with polarity $R_u$ and $R_v$, respectively.
Moreover, assume that $u$ is not connected to $v$ in the current instantiation of the network.
Let $p(u, v)$ be the probability that $u$ accepts a recommendation to connect to $v$.
We estimate $p(u, v)$ from training data.
Given a dataset of user interactions, we estimate $p(u, v)$ as the fraction 
$$N_{endorsed}(R_u, R_v) / N_{exposed}(R_u, R_v)$$
where $N_{exposed}(R_u, R_v)$ and $N_{endorsed}(R_u, R_v)$ are the number of times a user with polarity $R_v$ was {\it exposed to} or {\it endorsed}, respectively, content generated by a user of polarity $R_u$. $N_{exposed}(R_u, R_v)$ is computed by assuming that if $v$ follows $u$, $v$ is exposed to all content generated by $u$.
In practice, the polarity scores are bucketed together to avoid zero probabilities.
Experimental evaluation in Section~\ref{sec:exp_acceptance} shows that polarity scores learned this way predict the existence of an edge across datasets with good accuracy.

For a recommended edge $(u, v)$ from node $u$ to node $v$, with acceptance probability $p(u, v)$ and \rwc\ decrease $\delta \rwc_{u\rightarrow v}$, the {\it expected decrease} in \rwc\ when the edge is recommended individually is 
$$E(u, v) = p(u, v) \cdot \delta \rwc_{u\rightarrow v}.$$

Extending Algorithm~\ref{alg:algorithm1} to target the expected \rwc\ decrease can be done efficiently using Fagin's~\cite{fagin2003optimal} algorithm.
Specifically, we take as input two ranked lists of edges $(u, v)$, one ranked by decreasing $\delta \rwc_{u\rightarrow v}$ (as currently produced in the course of Algorithm~\ref{alg:algorithm1}) and another ranked by decreasing probability of acceptance $p(u, v)$.
Fagin's algorithm parses the two lists in parallel to find the edges that optimize the expected decrease $E(u, v)$.
We refer the interested reader to~\cite{fagin2003optimal} for details.



\section{Incremental computation of RWC}
\label{sec:recomputation}

The RWC score, as defined in Section~\ref{sec:rwc} can be computed using personalized pagerank compuations, usually implemented with power iterations. 
Since we are only interested in computing the incremental change in {\rwc} after adding an edge, we propose a way to quickly compute it.

Consider the transition probability matrix $P$.
After the addition of one (directed) edge from node $a$ to node $b$, only one column of $P$ is affected: the column that corresponds to the origin vertex ($a$) of the directed edge.
Let $q$ be the out degree of $a$. 
Specifically, before the addition of the edge, the $a^{th}$ column of the matrix has the following form.

\begin{equation}
P^T = \left[
\begin{array}{c}
... \\
... \\
\frac{1}{q}\ \frac{1}{q}\ \dots\ \frac{1}{q}\ 0\ 0\ \dots\ 0 \\
... \\
... 
\end{array}
\right]
\end{equation}

After adding the new outgoing edge from $a$, the transition probability matrix has the following form,
\begin{equation}
P^{'T} = \left[
\begin{array}{c}
... \\
... \\
\frac{1}{q+1}\ \frac{1}{q+1}\ \dots\ \frac{1}{q+1}\ \frac{1}{q+1}\ 0\ \dots\ 0 \\
... \\
... 
\end{array}
\right]
\end{equation}
with an additional $\frac{1}{q+1}$ at the $b^{th}$ index, and all other columns of the matrix are unchanged.

Define $\transpose{u} = \left[0\ 0\ 0\ 0\ \dots\ 1\ 0\ 0\ \dots\ 0\right]$
(the $a^{th}$ vector of the standard basis of $\mathbb{R}^n$).
Similarly, define $\transpose{v}$ as a column vector with a 1 at the $b^{th}$ position and 0 else where.

Define $\transpose{z}$: (i) If the outgoing vertex $a$ is not a dangling node, as: $\frac{1}{q+1} \cdot \left[\frac{1}{q}\ \frac{1}{q}\ \frac{1}{q}\ \frac{1}{q}\ \frac{1}{q}\ \dots\ \frac{1}{q}\ -1\ \dots\ 0\ 0\ 0\right]$ (i.e., $\frac{1}{q\cdot(q+1)}$ at all non zero neighbor indices, and $\frac{-1}{q+1}$ at the index of the incoming node). 
We can also say that $z_x$/$z_y$ is the column vector in $P_x$/$P_y$ corresponding to the outgoing node, multiplied by $\frac{1}{q+1}$ and a -1 at the index of the incoming node; and,
(ii) If the outgoing vertex is a dangling node, as: $e_x - v$ or $e_y - v$, depending on the side.

The updated transition probability matrix $P'$ is given by:
\begin{equation}
\label{eq:update_transition_matrix}
\begin{split}
P^{'} = P - z \cdot \transpose{u}.
\end{split}
\end{equation}

Let $M_x = I - \alpha \cdot P_x$ and $M_x^{'} = I - \alpha \cdot P_x^{'}$. 
Expanding the formula for $M_x'$, we get
\begin{equation}
\begin{split}
M_x^{'} = I - \alpha \cdot P_x^{'}  
 = I - \alpha P_x + \alpha z_x \transpose{u}
 = M_x + \alpha z_x \transpose{u}.
\end{split}
\end{equation}

Similarly for $M_y^{'}$, $M_y^{'}$ = $M_y\ +\ \alpha z_y\transpose{u}$.
As we can see, then, for any single edge addition, {\rwc} can be computed only using additional vectors that depends on the vertex that is affected.
Moreover, the inverse of $M_x^{'}$ (needed in Equation~(\ref{eq:rwc_simplified2})) can be computed efficiently using the Sherman Morisson formula.

\begin{lemma}[Sherman-Morrison Formula~\cite{golub2012matrix}]
Let $\mathbf{M}$ be a square $n \times n$ invertible matrix and $\mathbf{M}^{-1}$ its inverse.
Moreover, let $\mathbf{a}$ and $\mathbf{b}$ be any two column vectors of size $n$.
Then, the following equation holds
$$(\mathbf{M} + \mathbf{a}\mathbf{b}^{T})^{-1} = 
\mathbf{M}^{-1} - \mathbf{M}^{-1}\mathbf{a}\mathbf{b}^{T}\mathbf{M}^{-1} 
/ (1 + \mathbf{b}^{T}\mathbf{M}^{-1}\mathbf{a}).$$
\label{lemma:woodbury-simple}
\end{lemma}

Now, from Equation~(\ref{eq:rwc_simplified2}), the updated $RWC$, $RWC^{'}$ is,

$RWC^{'} = (1 - \alpha) \transpose{(c_x - c_y)} \cdot (M_x^{'-1}e_x - M_y^{'-1}e_y)$
, and the update in $RWC$ can be written as

\begin{equation}
\begin{split}
\small
\delta(RWC) = RWC^{'} - RWC \\
 = (1-\alpha) \transpose{(c_x - c_y)} \bigl((M_x^{'-1}e_x - M_x^{-1}e_x) \\
		 + (M_y^{-1}e_y - M_y^{'-1}e_y)\bigr)  \\
 = (1-\alpha) \transpose{(c_x - c_y)} \biggl( -\bigl(\frac{\alpha M_x^{-1}z_x \transpose{u} M_x^{-1}}{1+\alpha \transpose{u}M_x^{-1}z_x}\bigr)e_x \\
		 + \bigl(\frac{\alpha M_y^{-1}z_y \transpose{u} M_y^{-1}}{1+\alpha \transpose{u}M_y^{-1}z_y}\bigr)e_y \biggr) .
\end{split}
\label{eq:change_RWC}
\end{equation}
In light of Equation~(\ref{eq:change_RWC}), the costly inverse computation need not be performed in each iteration to compute the updated \rwc\ score.
When a new edge is added to the graph, we just compute the vectors $z_x$, $z_y$ and $u$ and use Equation~(\ref{eq:change_RWC}) to directly compute the incremental change in \rwc, instead of computing the new \rwc\ and taking the difference.
The matrix multiplication $M_*^{-1}z_* \transpose{u} M_*^{-1}$ can be computed efficiently by grouping the matrices as $(M_*^{-1}z_*) (\transpose{u} M_*^{-1})$.
As we see in Section~\ref{sec:timetaken}, this approach provides an order of magnitude speed up in the runtime of our algorithm.


\section{Experiments}
\label{sec:experiments}

In this section, we provide an evaluation of the two algorithms proposed in Section~\ref{sec:algorithms}. We use the acronym \rov\ ({\it \underline{r}ecommend \underline{o}pposing \underline{v}iew}) to refer to Algorithm~\ref{alg:algorithm1}; and \rovap\ ({\it \underline{r}ecommend \underline{o}pposing \underline{v}iew - with \underline{a}cceptance \underline{p}roba-bility}) to refer to its variation that also considers edge acceptance probabilities.

\subsection{Datasets}

We use Twitter datasets on known controversial issues.
The datasets have also been used in previous studies~\cite{garimella2016quantifying,lu2015biaswatch}.
Dataset statistics are shown in Table~\ref{tab:datasets}.
Eight of the datasets consist of tweets collected by tracking single hashtags over a small period of time.
The remaining two datasets ({\it obamacare}, {\it guncontrol}) 
consist of tweets collected via the 
twitter streaming API\footnote{\url{https://dev.twitter.com/streaming/public}} 
by tracking the corresponding keywords for two years. 
We process the datasets and construct {\em retweet graphs}.
We remark that even though all our datasets are from Twitter, 
our work can be applied on any graph with a clustered structure separating the sides of a controversy.

\begin{table}[t]
\caption{\label{tab:datasets}Datasets statistics: hashtag
used to collect dataset, number of tweets, size of retweet graph.}
\centering
\small
\begin{tabular}{l r r r}
\toprule
\multicolumn{1}{l}{\bf Dataset} & \multicolumn{1}{c}{\bf \# Tweets} & \multicolumn{2}{c}{\bf Retweet graph} \\
\cmidrule(lr){3-4}
 & & \multicolumn{1}{c}{$|V|$} & \multicolumn{1}{c}{$|E|$} \\
\midrule
\#beefban & \num{84543} & \num{1610} & \num{1978} \\ 
\#nemtsov & \num{183477} & \num{6546} & \num{10172} \\ 
\#netanyahuspeech & \num{254623} & \num{9434} & \num{14476} \\
\#russia\_march & \num{118629} & \num{2134} & \num{2951} \\
\#indiasdaughter & \num{167704} & \num{3659} & \num{4323} \\
\#baltimoreriots & \num{218157} & \num{3902} & \num{4505} \\
\#indiana & \num{116379} & \num{2467} & \num{3143} \\
\#ukraine & \num{287438} & \num{5495} & \num{9452} \\
obamacare & \num{123320} & \num{3132} & \num{3241} \\
guncontrol & \num{117679} & \num{2633} & \num{2672} \\
\bottomrule
\end{tabular}
\vspace{-\baselineskip}
\end{table}

\subsection{Comparison with other link prediction and recommendation systems}
\label{sec:exp_acceptance}

In this section, 
we evaluate the choice of using node polarity scores 
for predicting edge acceptance (Section~\ref{sec:acceptance}). 
To perform this evaluation 
by comparing our approach with other state-of-the-art link-prediction algorithms, 
which are listed in Table~\ref{tab:link_rec_comparison}. 

\begin{table}[t]
\centering
\small
\caption{Algorithms explored for link prediction.}
\label{tab:link_rec_comparison}
\begin{tabular}{llr}
\toprule
\textbf{Algorithm} & \textbf{Summary} & \multicolumn{1}{c}{\textbf{AUC}} \\
\midrule
Node polarity & Link recommendation based on & 0.79
\\ &  node polarity &  \\
Adamic-Adar~\cite{adamic2003friends}  & Link prediction based on number & 0.60 \\ 
&  of common neighbors  &  \\
Reliability~\cite{zhou2009predicting} & Block stochastic model & 0.66 \\
RAI~\cite{soundarajan2012using} & Using community detection to & 0.60 \\ 
 & improve link prediction &  \\
SLIM~\cite{ning2011slim}  & Collaborative filtering & 0.71 \\ 
& recommendation &  \\
FISM~\cite{kabbur2013fism} & Content-based recommendation & 0.66 \\ 
\bottomrule
\end{tabular}
\vspace{-\baselineskip}
\end{table}

Following Section~\ref{sec:acceptance}, 
to estimate acceptance probabilities as a function of user polarity, 
we first bucket the user polarity scores into 10 equally sized buckets, from -1 to +1.
Then, 
we estimate acceptance probabilities $p(u, v)$ 
separately for each bucket combination of users $u$ and $v$.
We train a model and cross-validate across all datasets. 
The median AUC is 0.79, 
indicating that 
endorsement graphs across different datasets have similar edge-formation criteria.

We compare our approach with existing link-recommendation methods.
The implementations are obtained from Librec~\cite{guo2015librec}. 
The results are reported in Table~\ref{tab:link_rec_comparison}.
%
%
As we can see, 
our approach of using node-polarity scores for predicting links 
works as well as the best link-recommendation algorithm. 
Note that the objective here is not to propose yet another link-recommendation algorithm,
nor to claim that our method works better than other approaches in general. 
The objective is to validate the use of node polarities 
for creating a model for edge-acceptance probabilities.

\subsection{Comparison with other related approaches}


As mentioned earlier, 
this is the first paper that addresses the problem of selecting edges to add for decreasing controversy.
However, there have been other methods that consider adding edges for improving
other structural graph properties. 
In this section, 
we compare our approach with three such recent methods:
($i$) NetGel~\cite{tong2012gelling}, which maximizes the largest eigenvalue; 
($ii$) MioBi~\cite{chan2014make}, which maximizes the average eigenvalue; and 
($iii$) Shortcut~\cite{parotisidis2015selecting}, which minimizes the average shortest path. 
We also experiment with the simple greedy version of our approach, 
which does not use the heuristic proposed in Section~\ref{sec:algorithms}, 
but considers all possible 
edges.

\begin{figure*}[t]
\begin{minipage}{.19\linewidth}
\centering
\subfloat[]{\label{}\includegraphics[width=\textwidth, height=\textwidth]{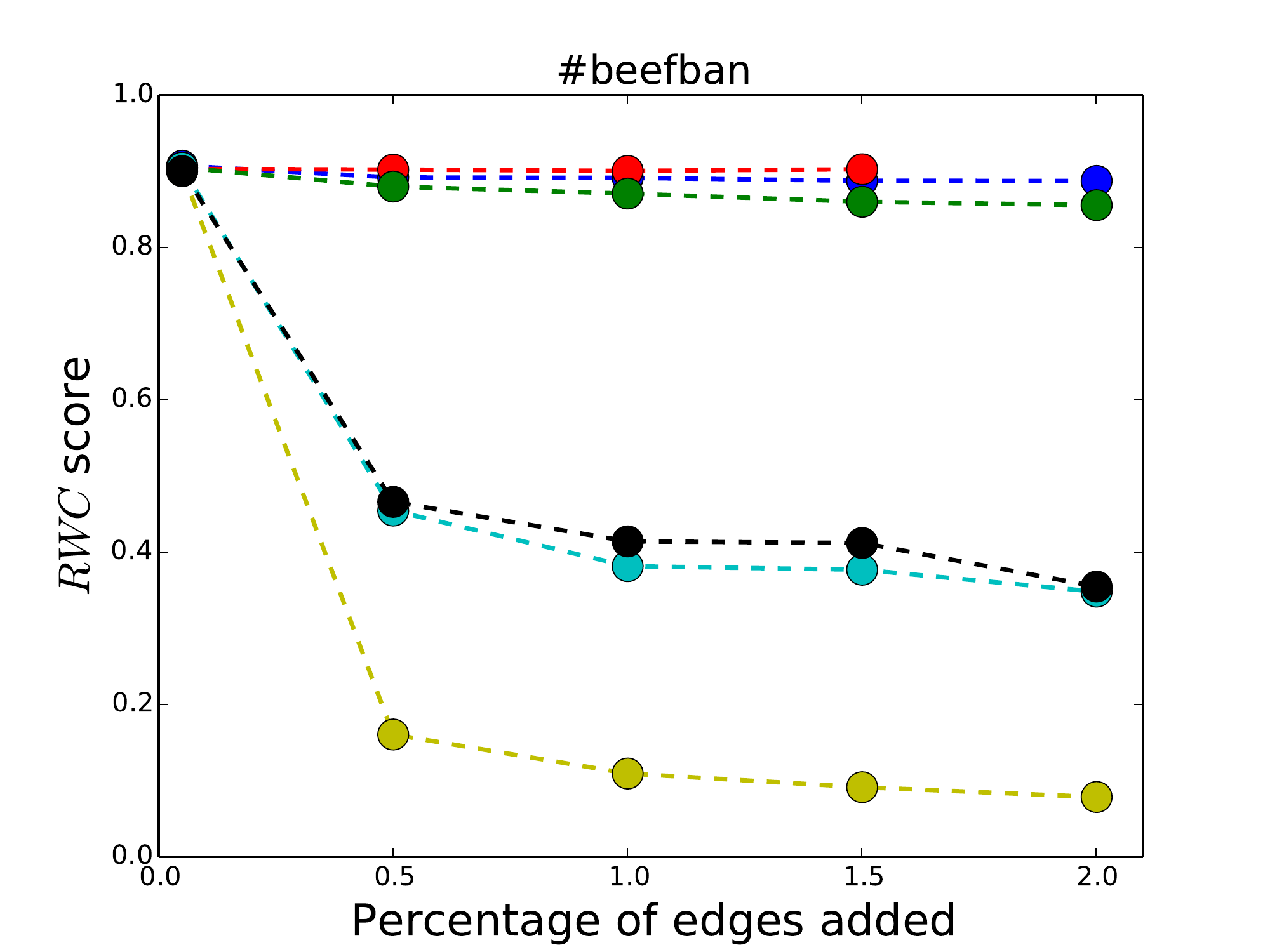}}
\end{minipage}%
\begin{minipage}{.19\linewidth}
\centering
\subfloat[]{\label{}\includegraphics[width=\textwidth, height=\textwidth]{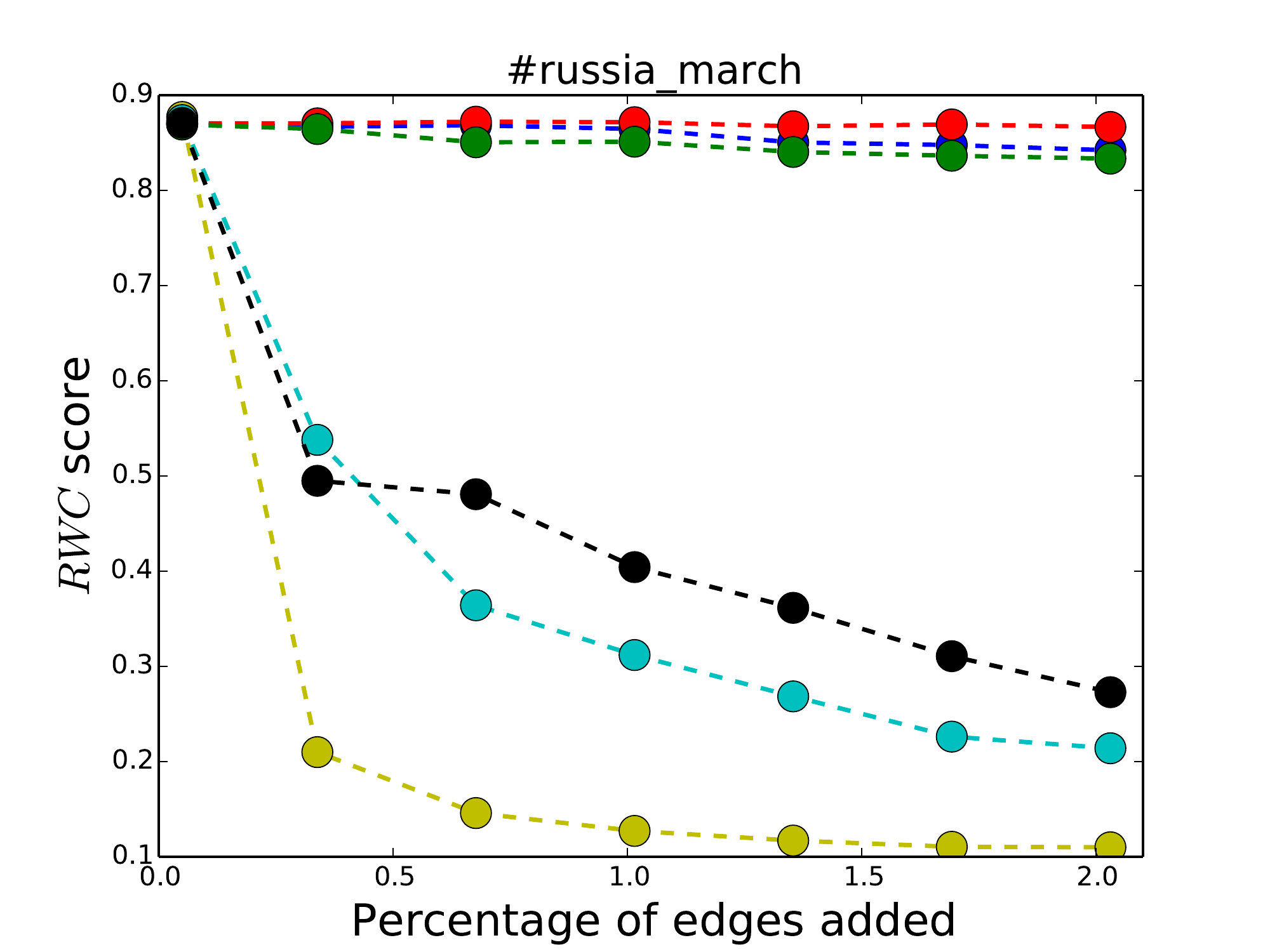}}
\end{minipage}
\begin{minipage}{.19\linewidth}
\centering
\subfloat[]{\label{}\includegraphics[width=\textwidth, height=\textwidth]{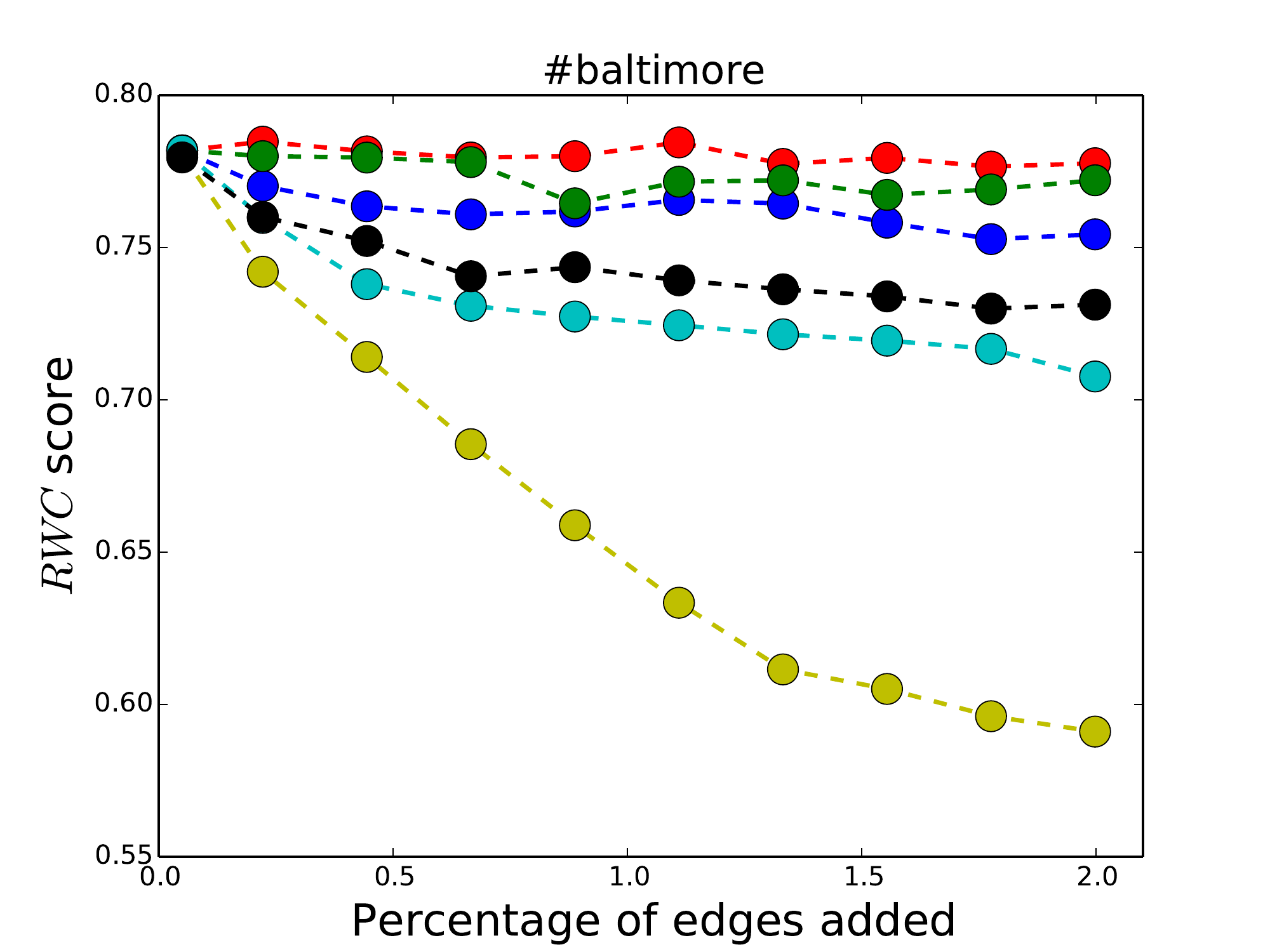}}
\end{minipage}
\begin{minipage}{.19\linewidth}
\centering
\subfloat[]{\label{}\includegraphics[width=\textwidth, height=\textwidth]{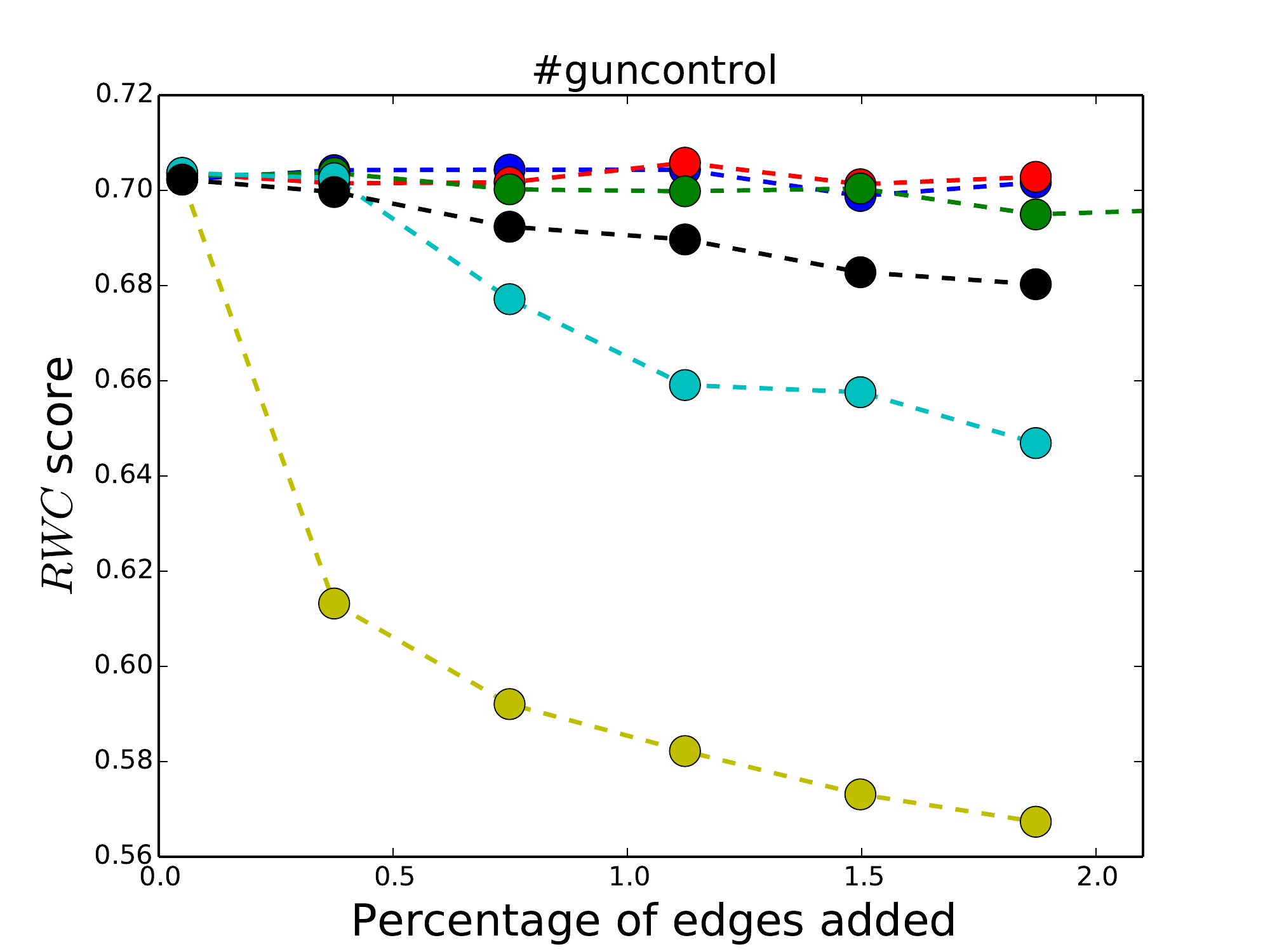}}
\end{minipage}
\begin{minipage}{.19\linewidth}
\centering
\subfloat[]{\label{}\includegraphics[width=\textwidth, height=\textwidth]{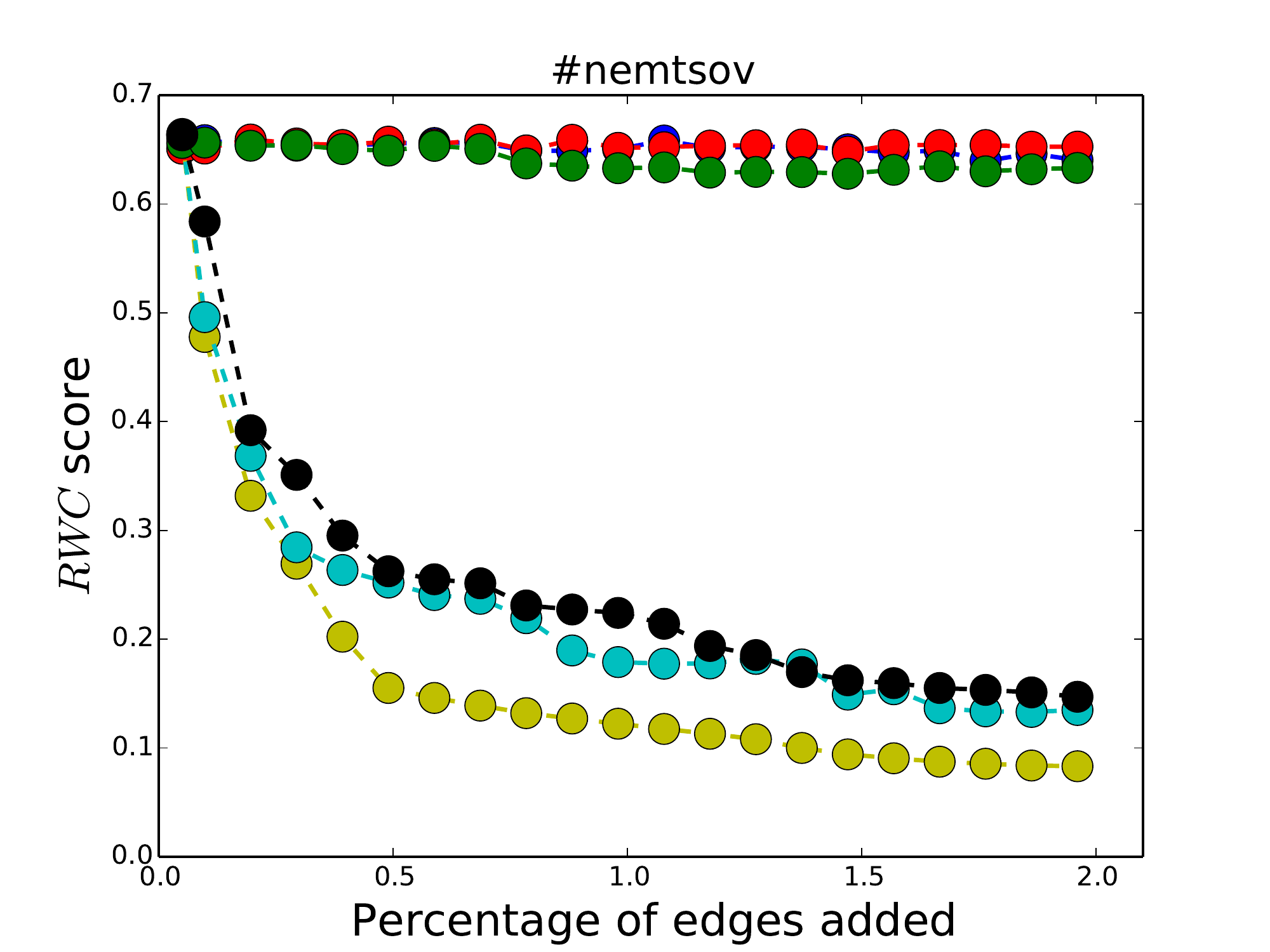}}
\end{minipage}\par\medskip
\begin{minipage}{.19\linewidth}
\centering
\subfloat[]{\label{}\includegraphics[width=\textwidth, height=\textwidth]{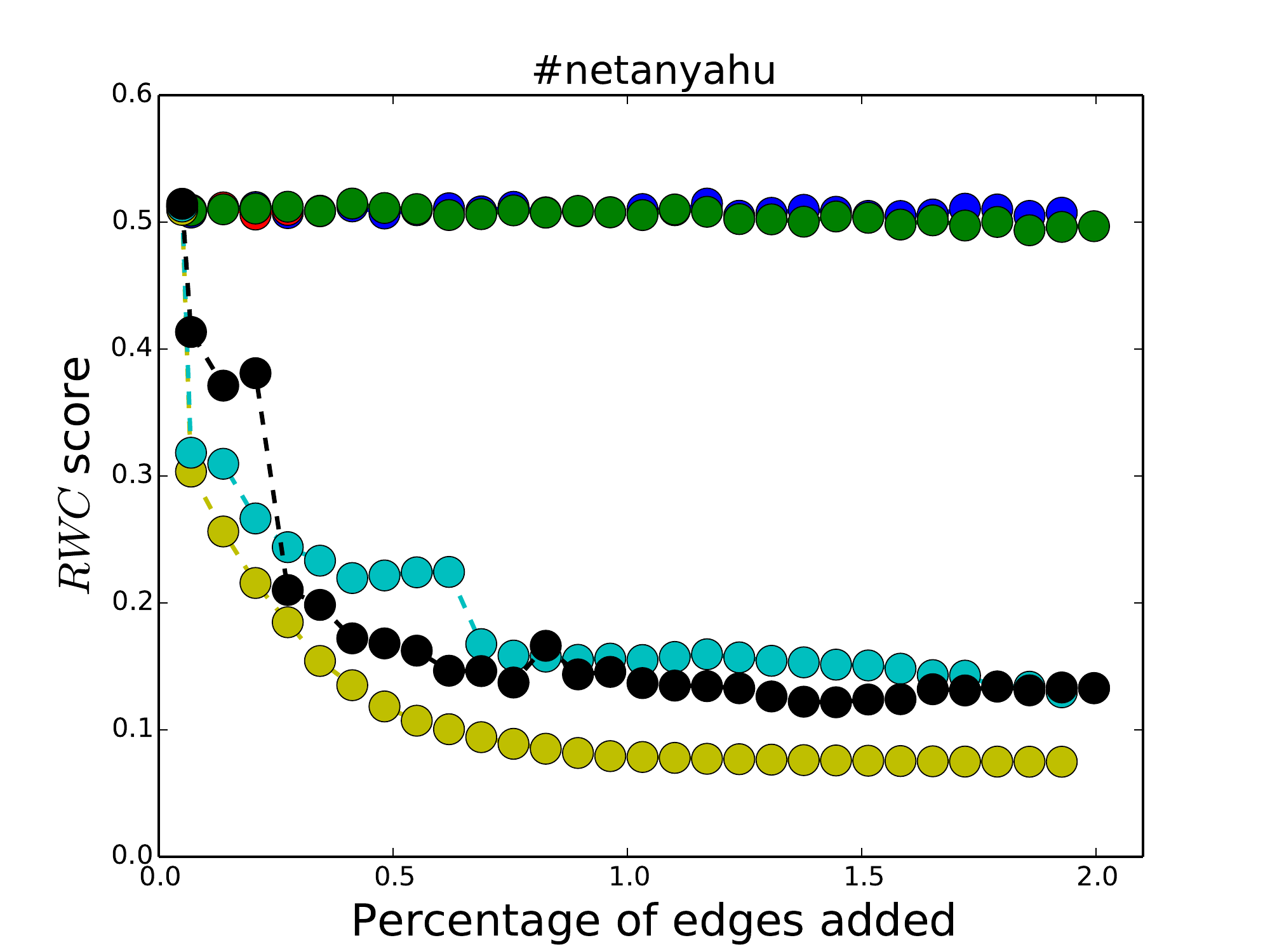}}
\end{minipage}
\begin{minipage}{.19\linewidth}
\centering
\subfloat[]{\label{}\includegraphics[width=\textwidth, height=\textwidth]{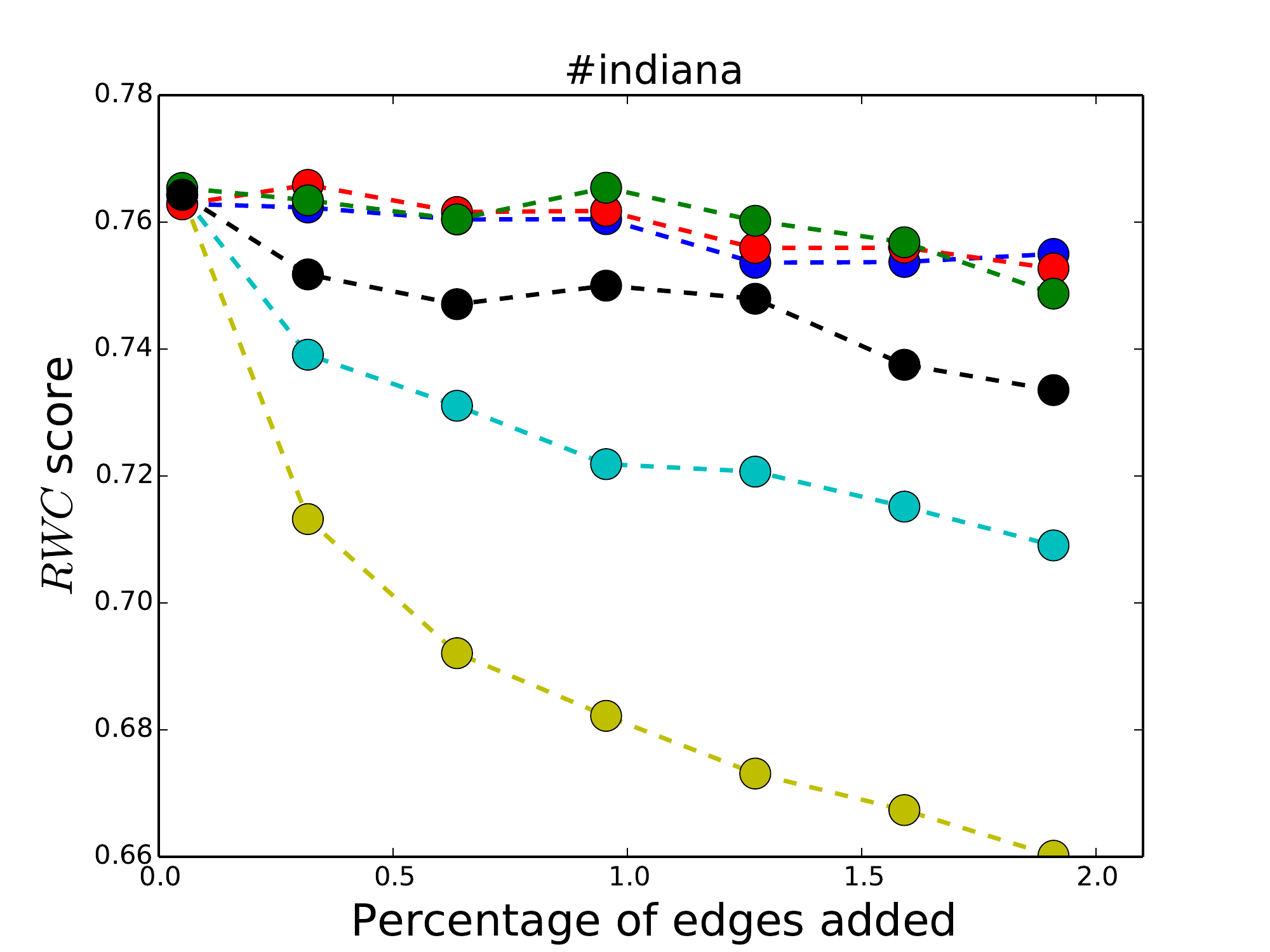}}
\end{minipage}
\begin{minipage}{.19\linewidth}
\centering
\subfloat[]{\label{}\includegraphics[width=\textwidth, height=\textwidth]{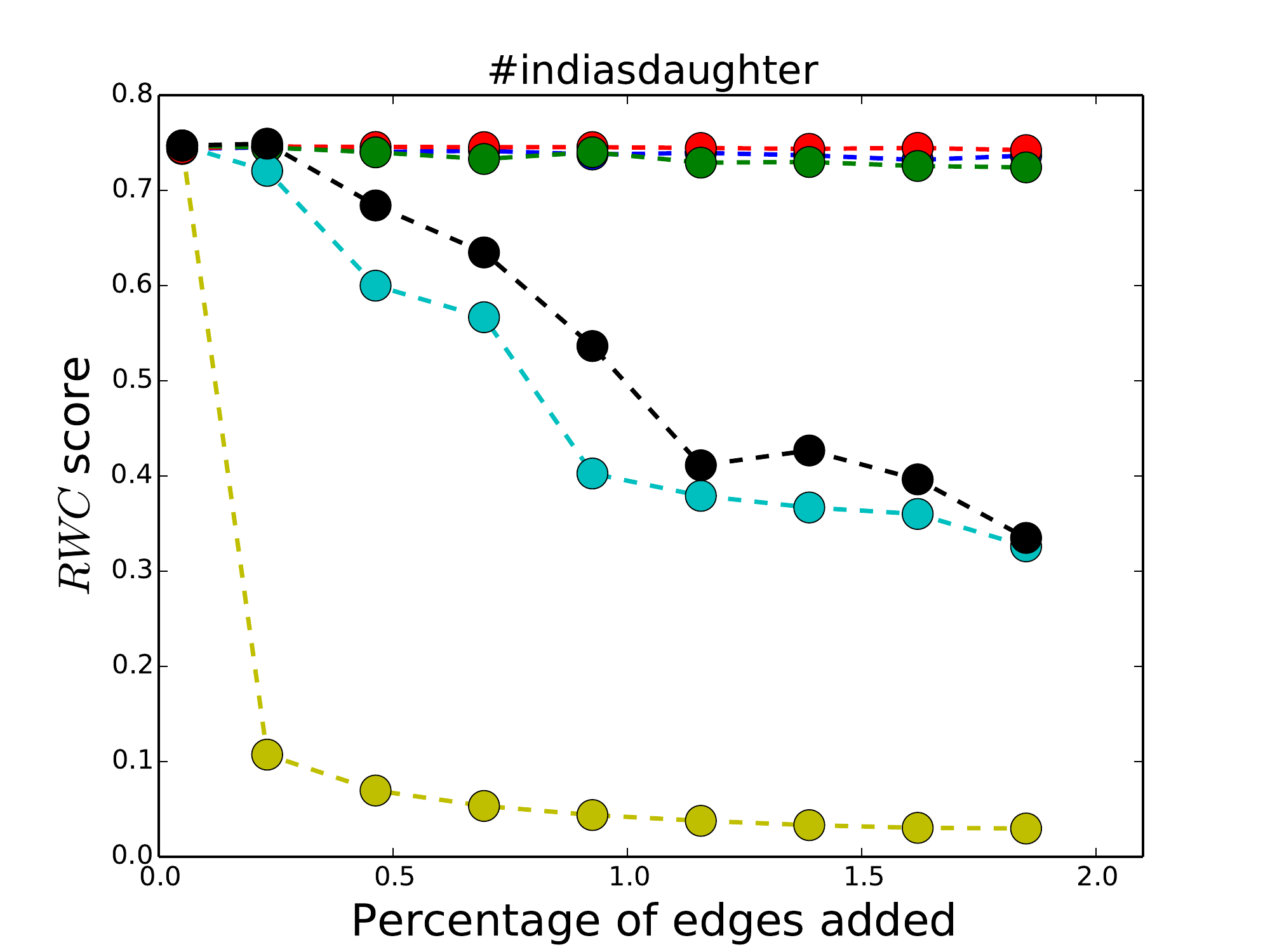}}
\end{minipage}
\begin{minipage}{.19\linewidth}
\centering
\subfloat[]{\label{}\includegraphics[width=\textwidth, height=\textwidth]{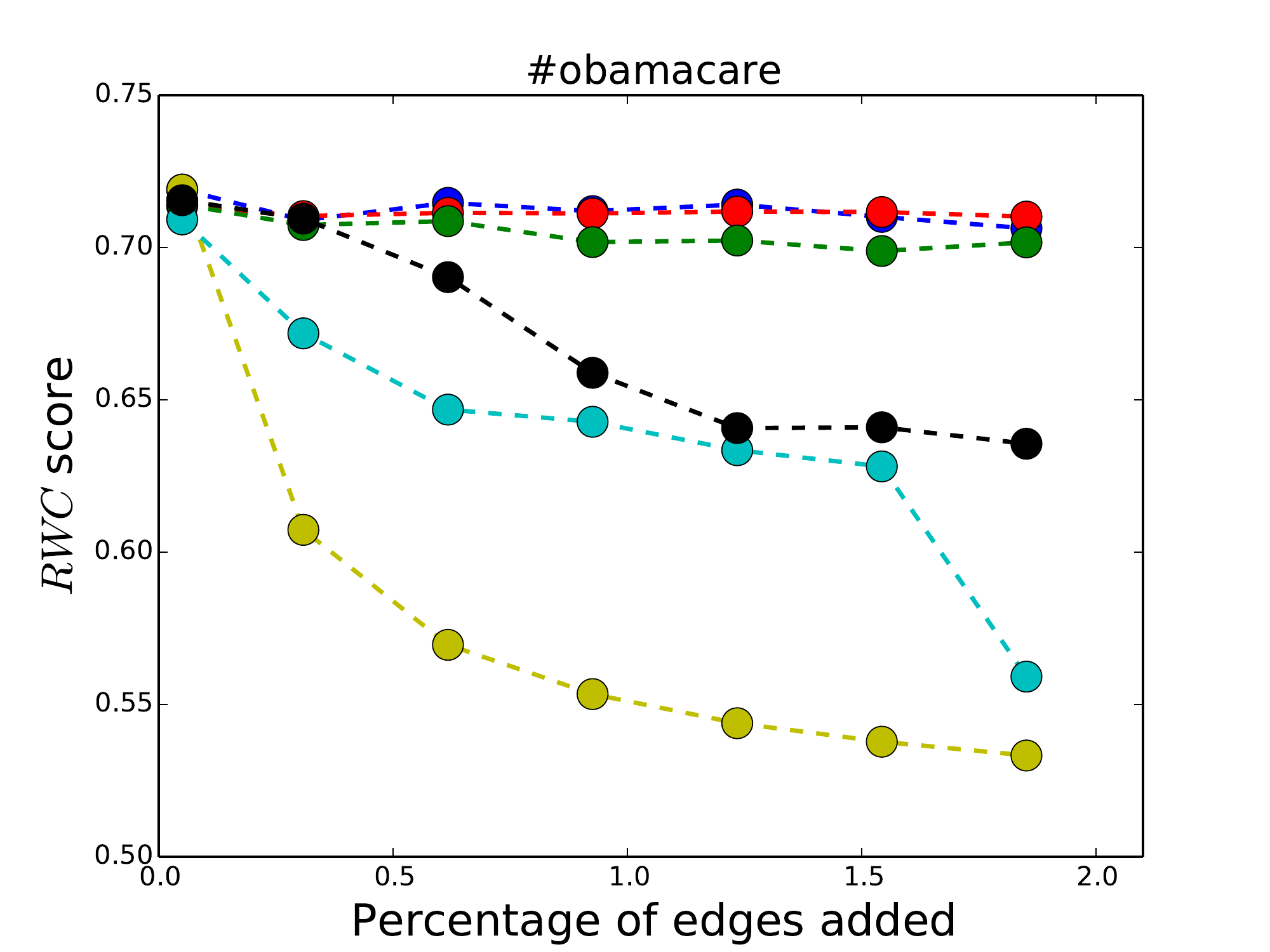}}
\end{minipage}
\begin{minipage}{.19\linewidth}
\centering
\subfloat[]{\label{}\includegraphics[width=\textwidth, height=\textwidth]{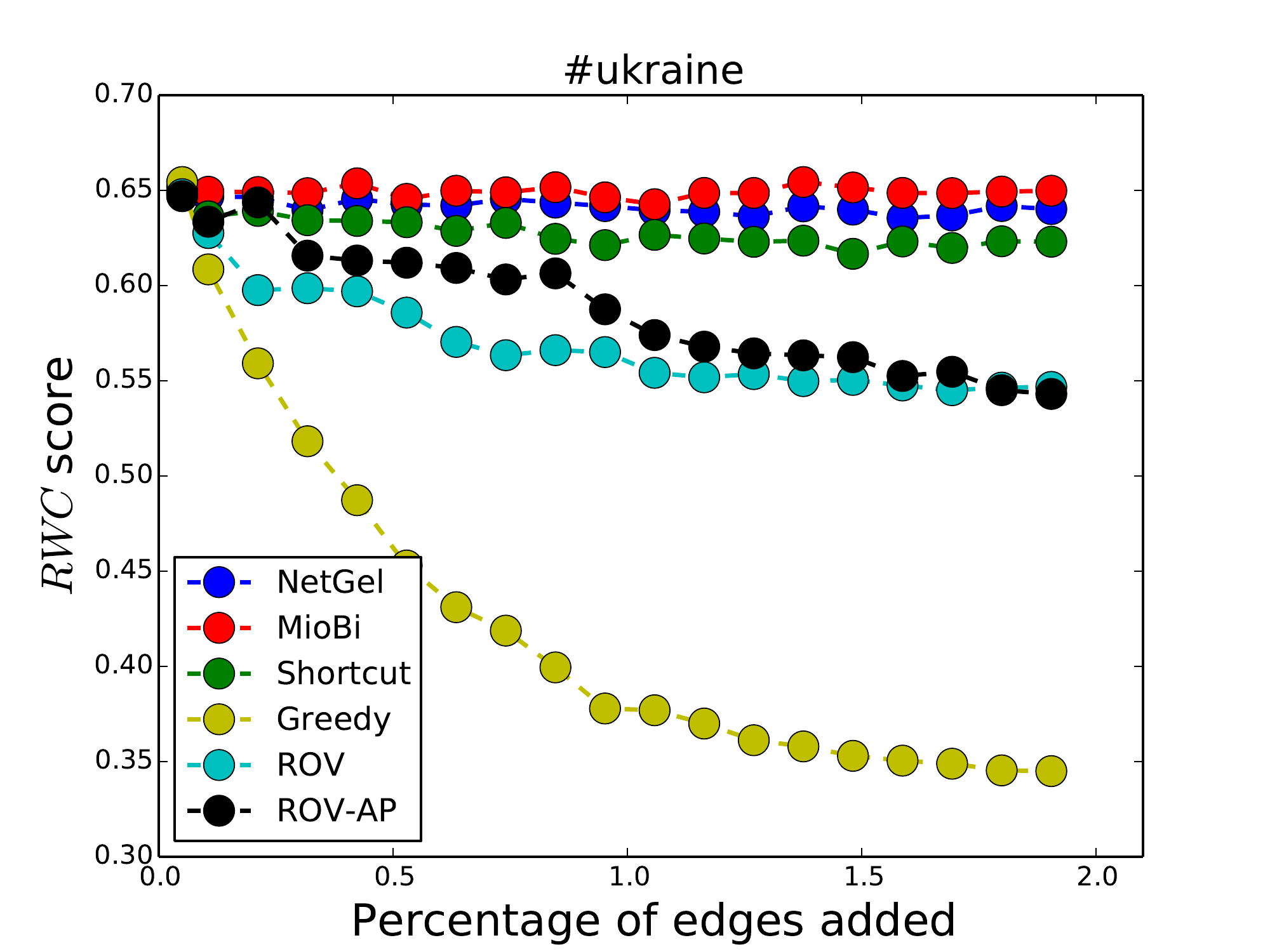}}
\end{minipage}\par\medskip
\caption{Comparison 
of the proposed methods ({\rov} and {\rovap}) 
with related approaches (NetGel, MioBi, Shortcut) 
for 2\% of the total edges added. 
The Greedy algorithm considers all possible edges.}
\label{fig:comparison_related}
\vspace{-\baselineskip}
\end{figure*}

The results are shown in Figure~\ref{fig:comparison_related}. 
As expected, the greedy brute-force algorithm
performs the best. 
Our algorithm, {\rov}, which considers only a small fraction of possible edges, 
performs quite well, and in some cases, is on par with the greedy.
The version of our algorithm with edge acceptance probabilities, {\rovap}, comes next. 
It is worth noting that even though the choice of edges for {\rovap} is based on a different criterion, 
the performance of the algorithm in terms of the {\rwc} score is not impacted much.
On the other hand, 
as we will see in Section~\ref{sec:casestudy}, 
using edge acceptance probabilities improves significantly the real world applicability of our approach.

The other methods (NetGet, MioBi and Shortcut) 
do not perform particularly well. 
This is expected, as those methods are not designed to optimize our objective function.
Overall, our results demonstrate the need for a specialized method to reduce controversy.

\subsection{Edge-addition strategies}

\begin{figure*}[t]
\begin{minipage}{.19\linewidth}
\centering
\subfloat[]{\label{}\includegraphics[width=\textwidth, height=\textwidth]{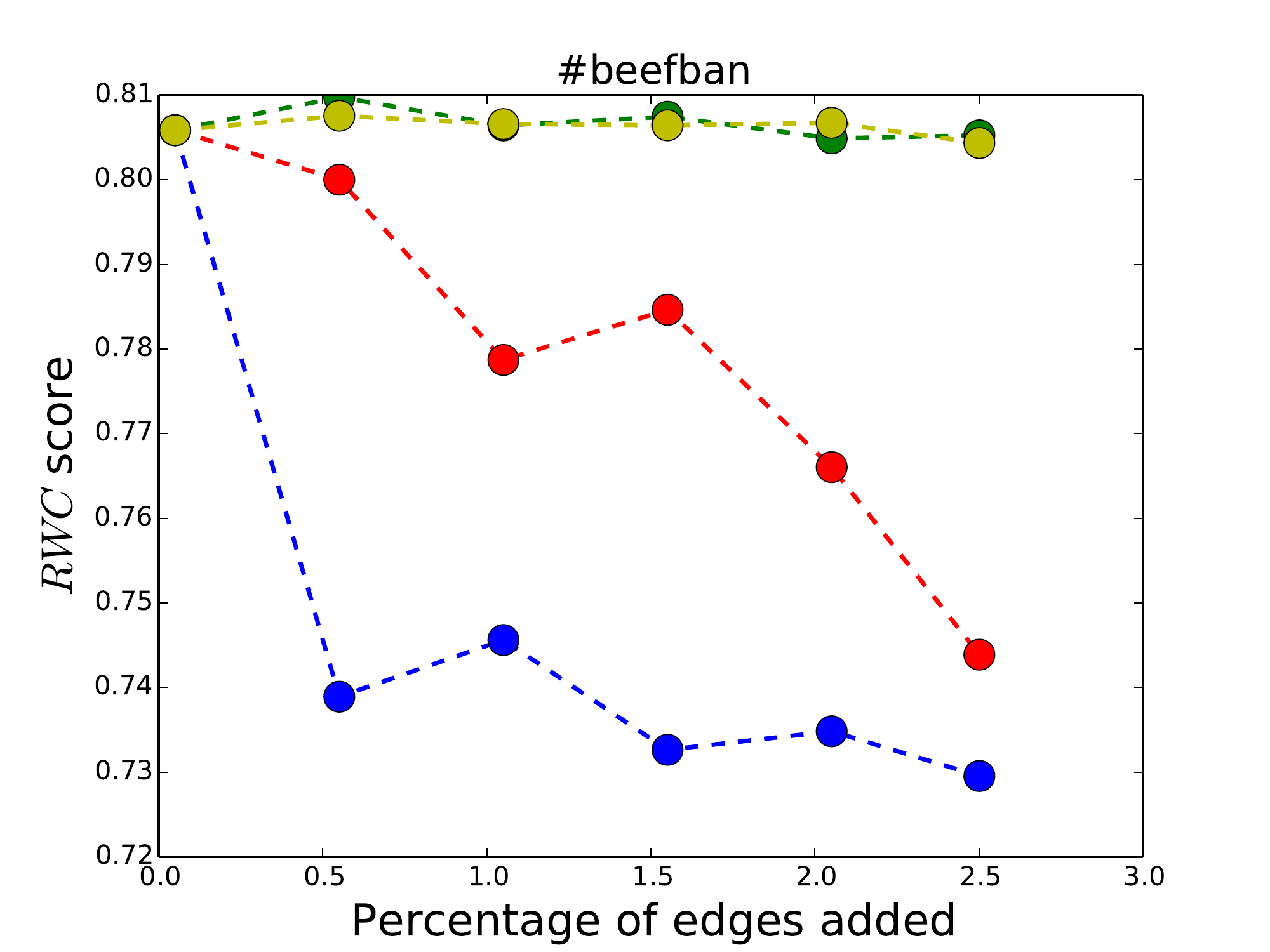}}
\end{minipage}%
\begin{minipage}{.19\linewidth}
\centering
\subfloat[]{\label{}\includegraphics[width=\textwidth, height=\textwidth]{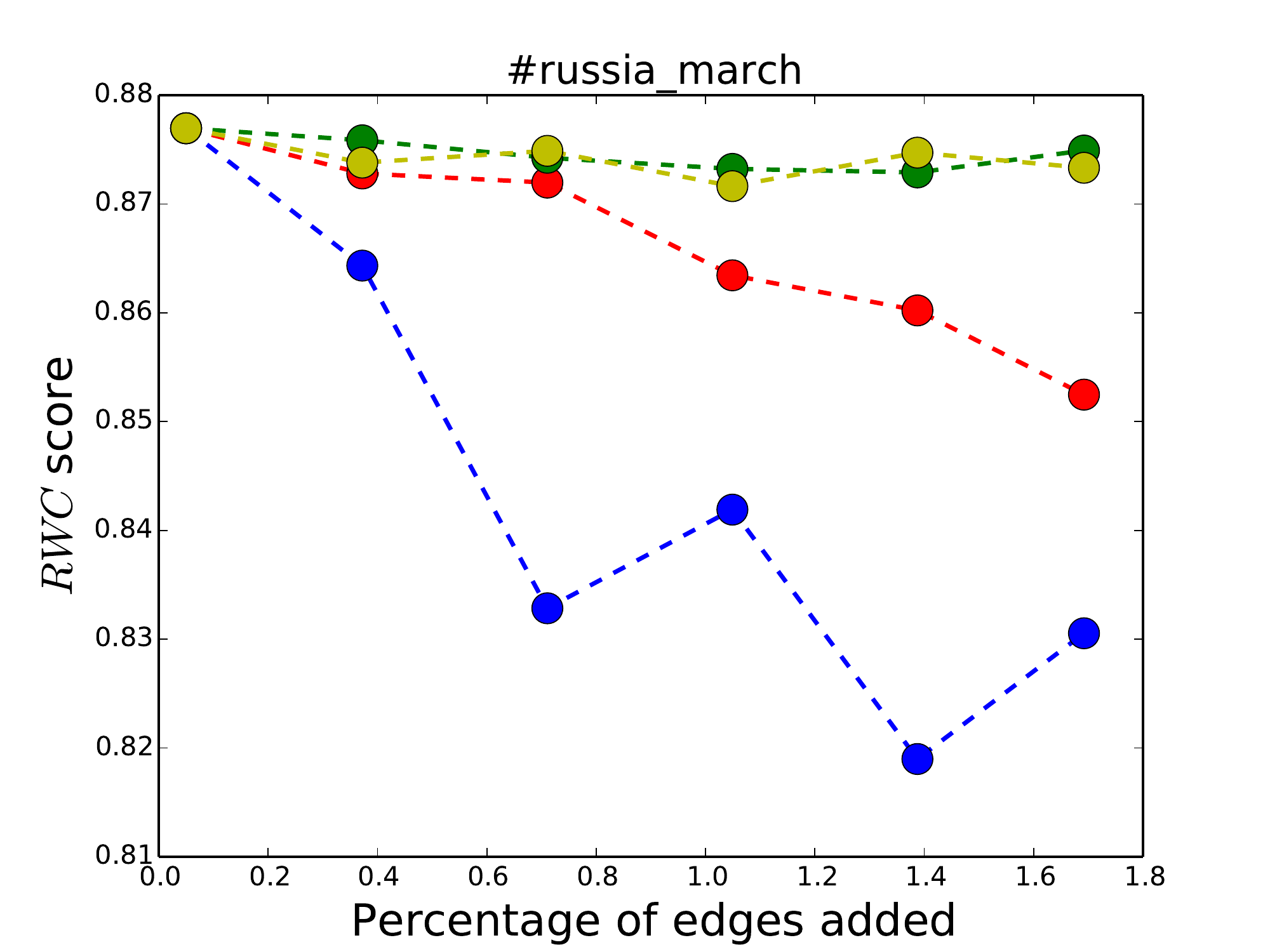}}
\end{minipage}
\begin{minipage}{.19\linewidth}
\centering
\subfloat[]{\label{}\includegraphics[width=\textwidth, height=\textwidth]{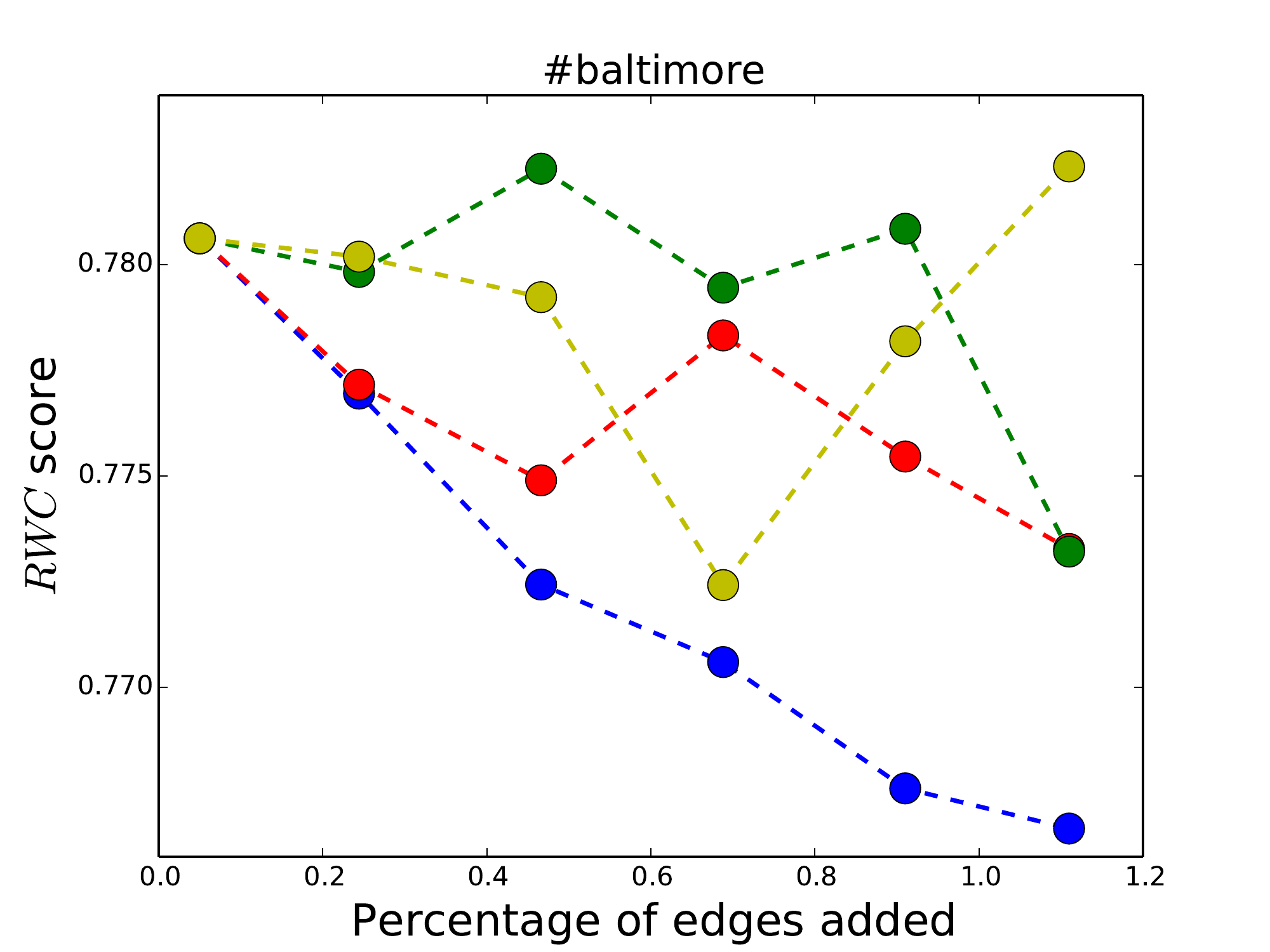}}
\end{minipage}
\begin{minipage}{.19\linewidth}
\centering
\subfloat[]{\label{}\includegraphics[width=\textwidth, height=\textwidth]{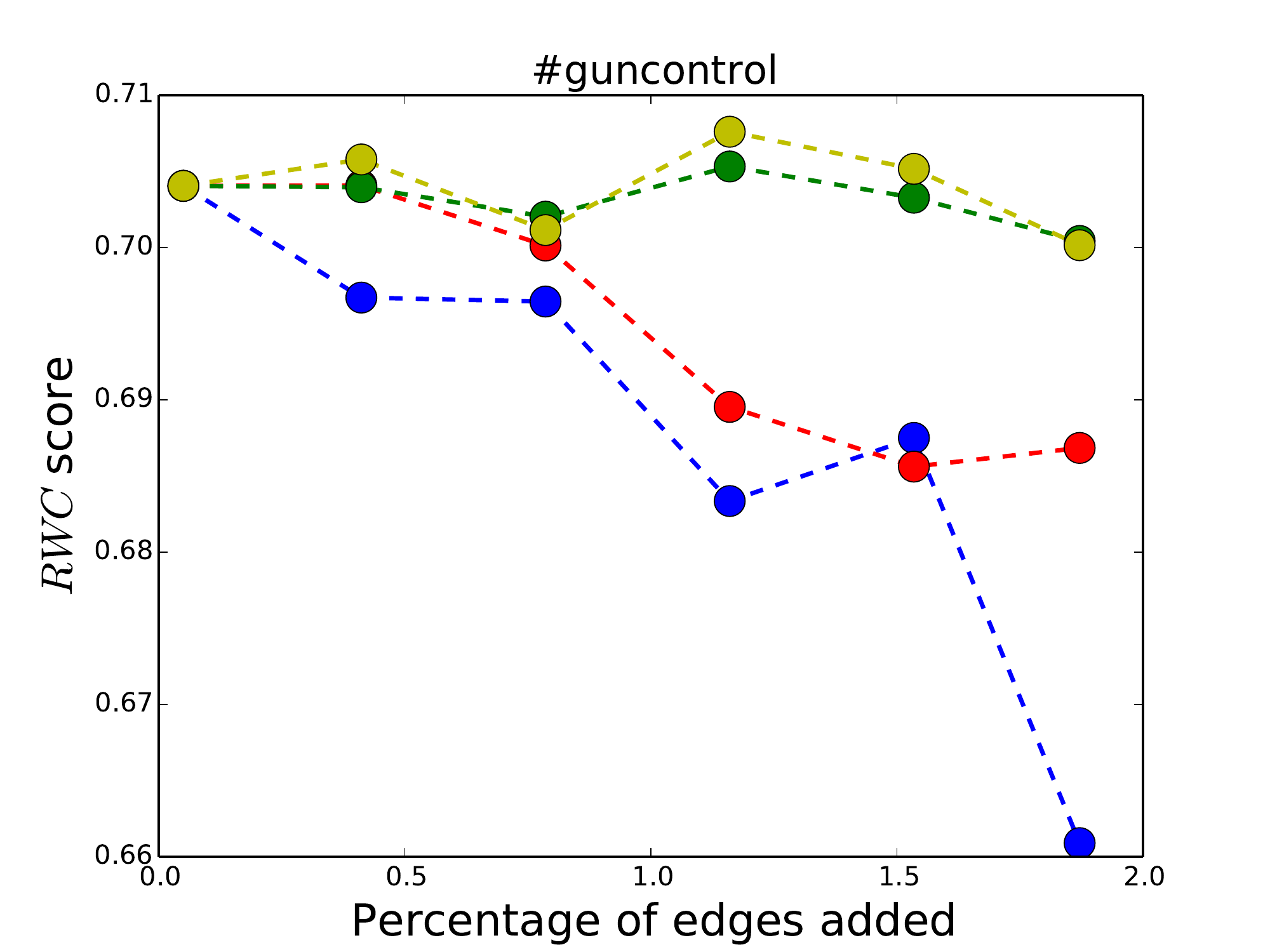}}
\end{minipage}
\begin{minipage}{.19\linewidth}
\centering
\subfloat[]{\label{}\includegraphics[width=\textwidth, height=\textwidth]{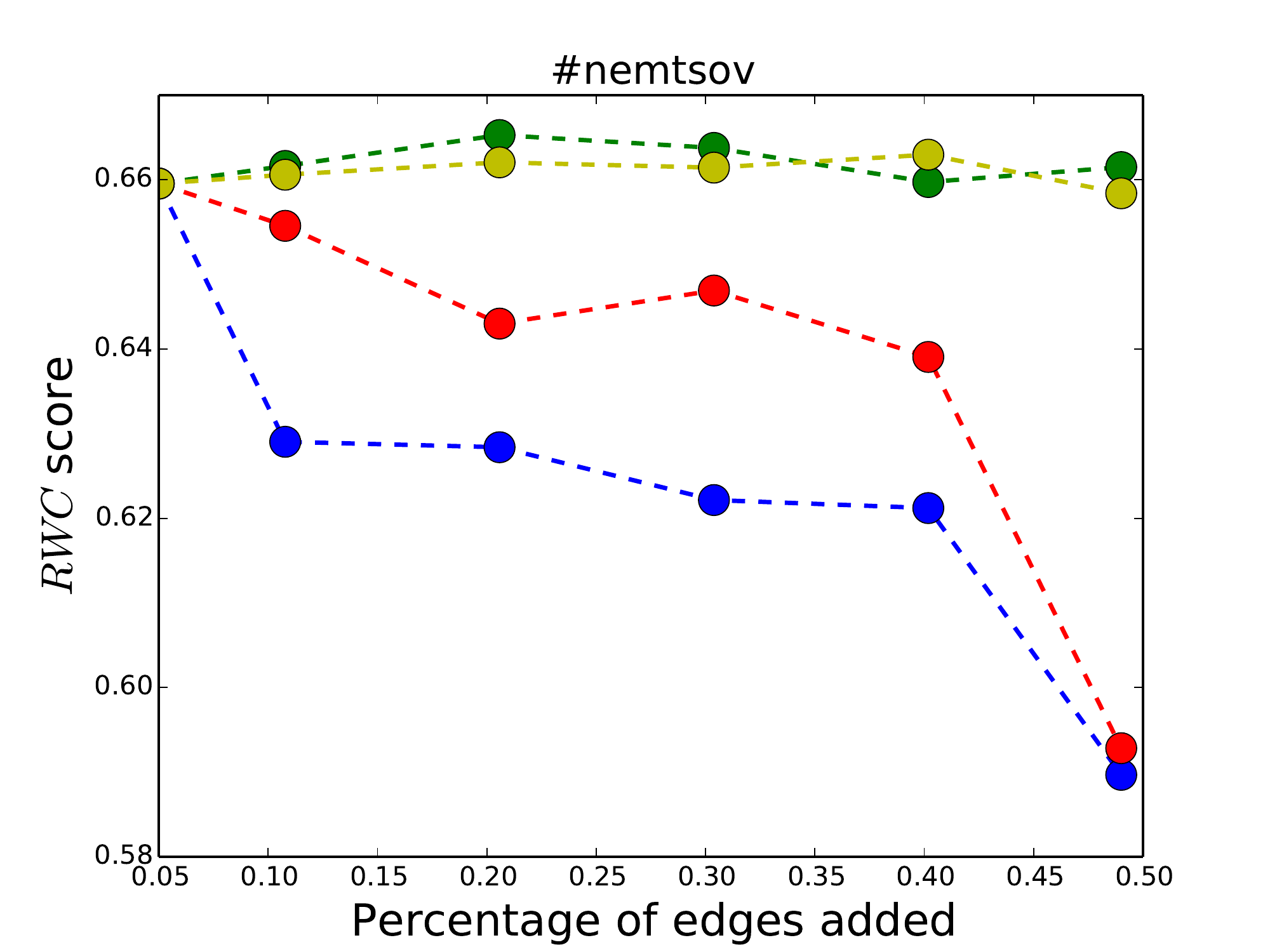}}
\end{minipage}\par\medskip
\begin{minipage}{.19\linewidth}
\centering
\subfloat[]{\label{}\includegraphics[width=\textwidth, height=\textwidth]{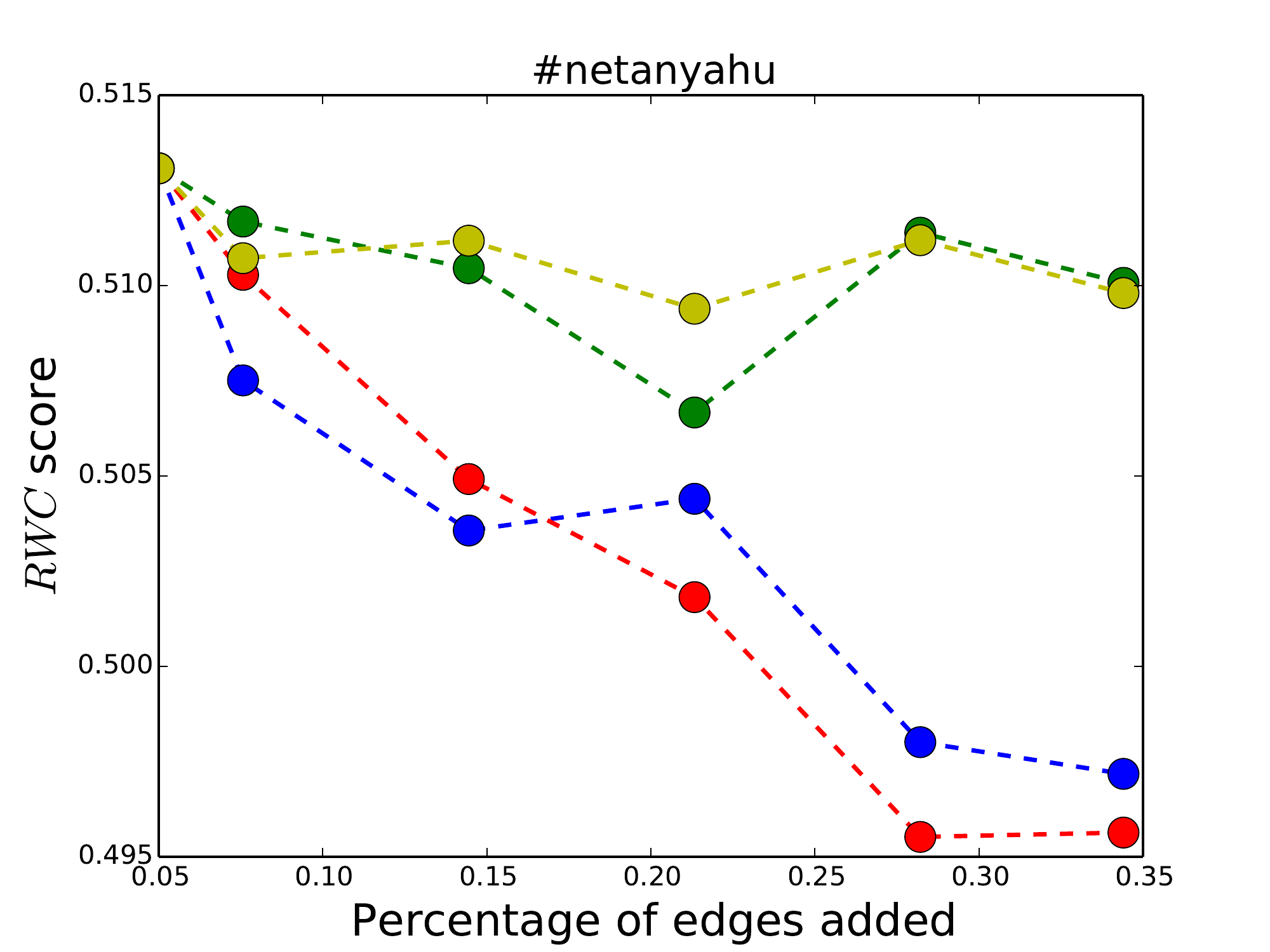}}
\end{minipage}
\begin{minipage}{.19\linewidth}
\centering
\subfloat[]{\label{}\includegraphics[width=\textwidth, height=\textwidth]{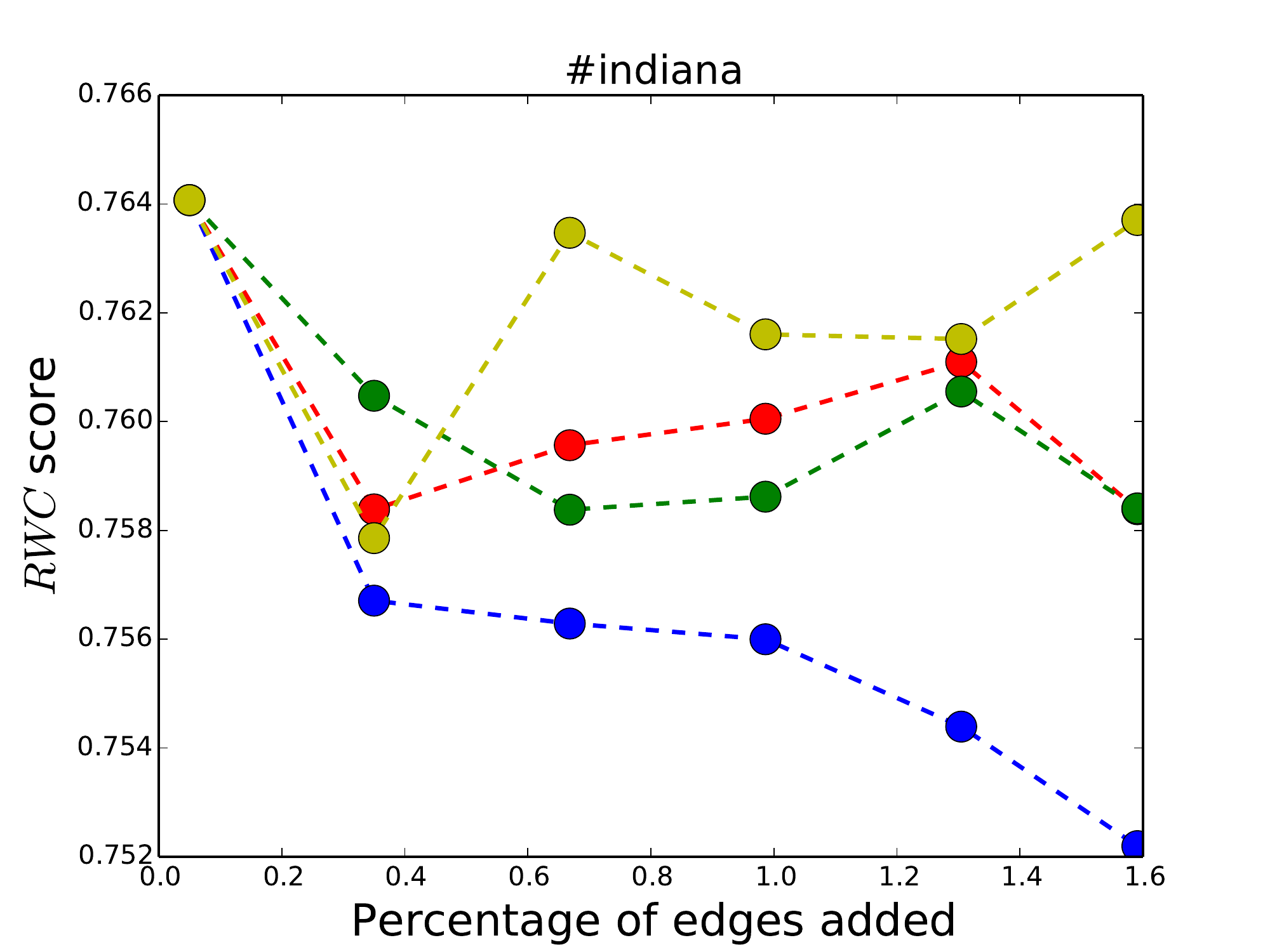}}
\end{minipage}
\begin{minipage}{.19\linewidth}
\centering
\subfloat[]{\label{}\includegraphics[width=\textwidth, height=\textwidth]{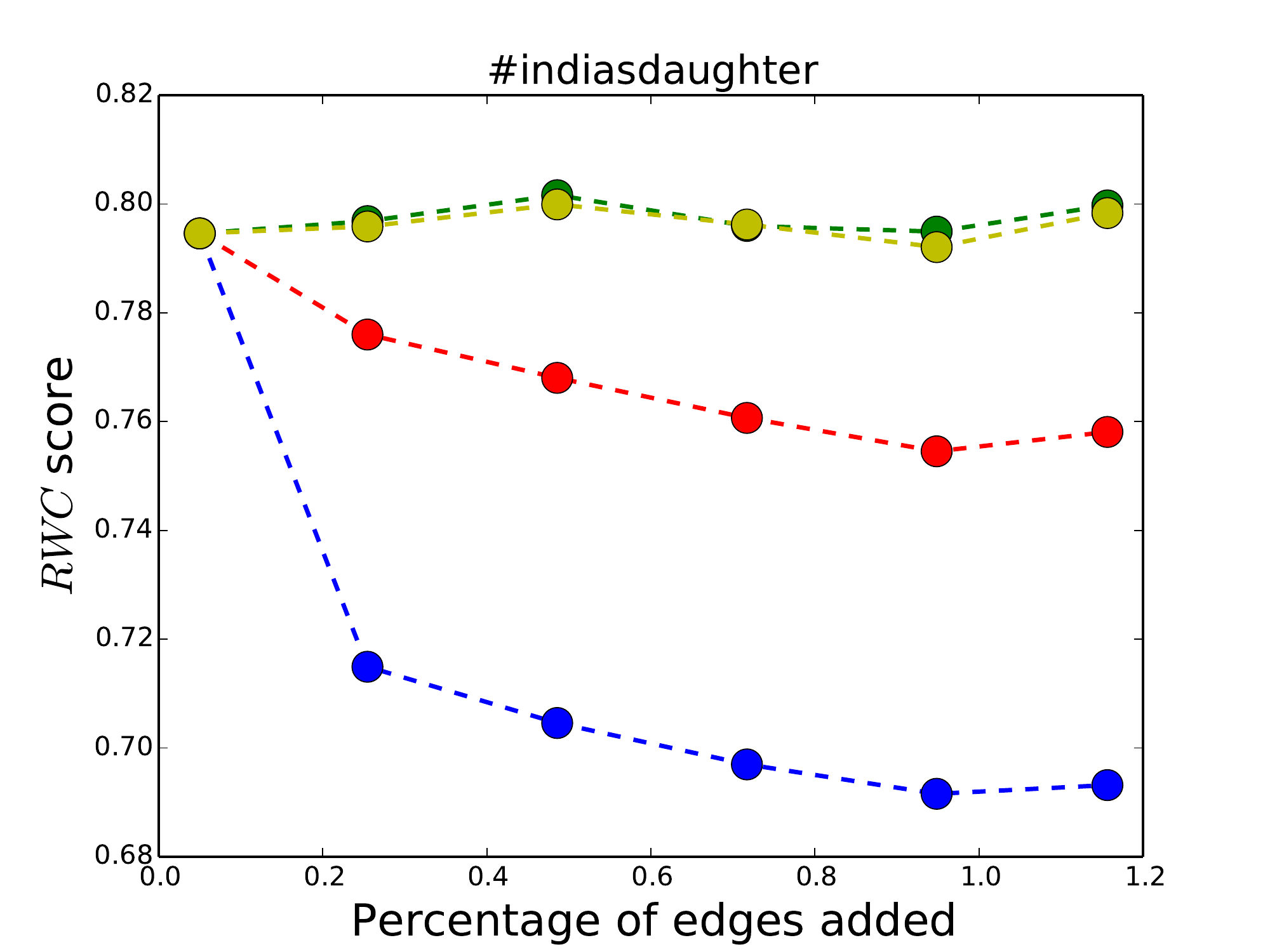}}
\end{minipage}
\begin{minipage}{.19\linewidth}
\centering
\subfloat[]{\label{}\includegraphics[width=\textwidth, height=\textwidth]{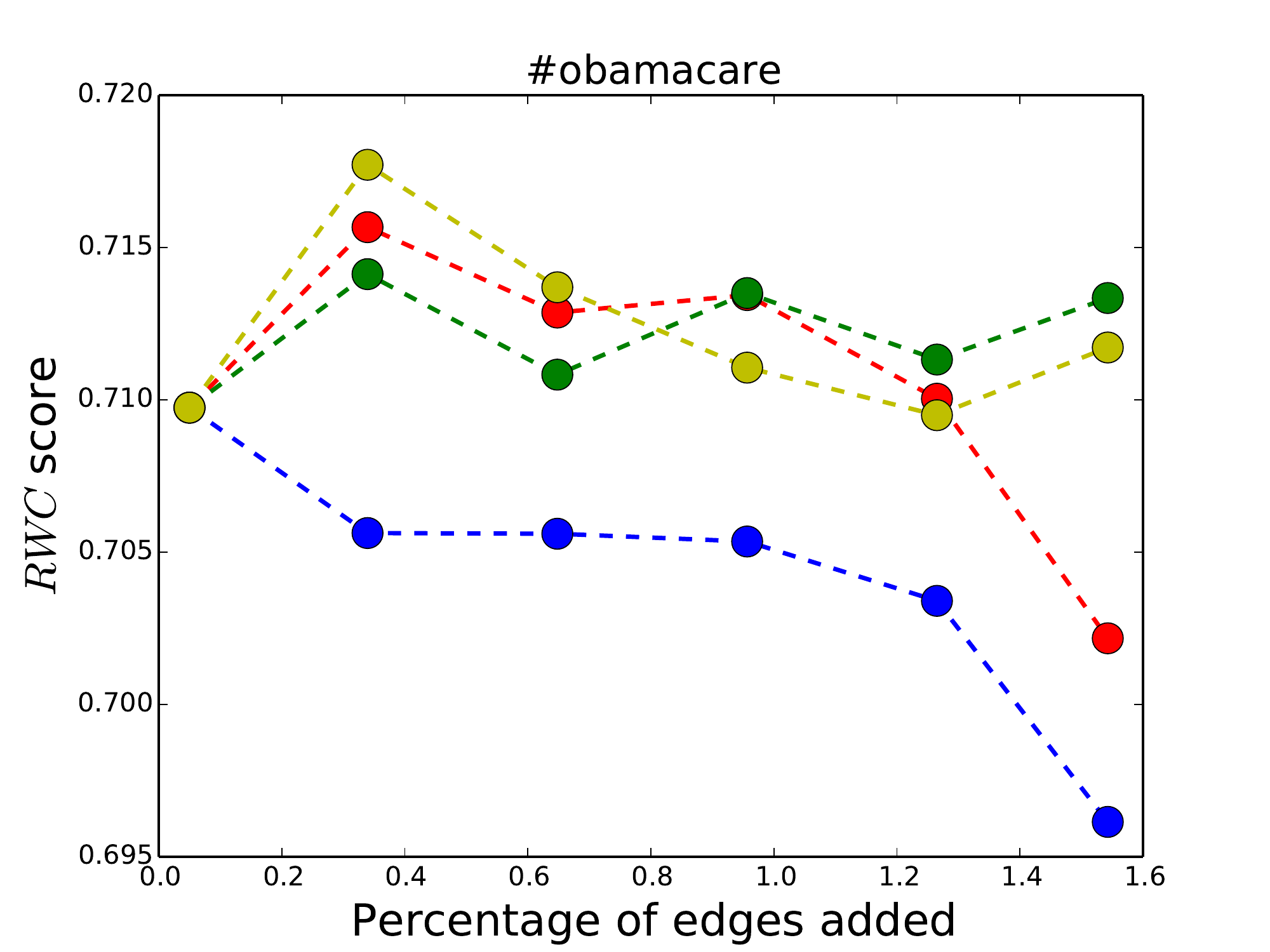}}
\end{minipage}
\begin{minipage}{.19\linewidth}
\centering
\subfloat[]{\label{}\includegraphics[width=\textwidth, height=\textwidth]{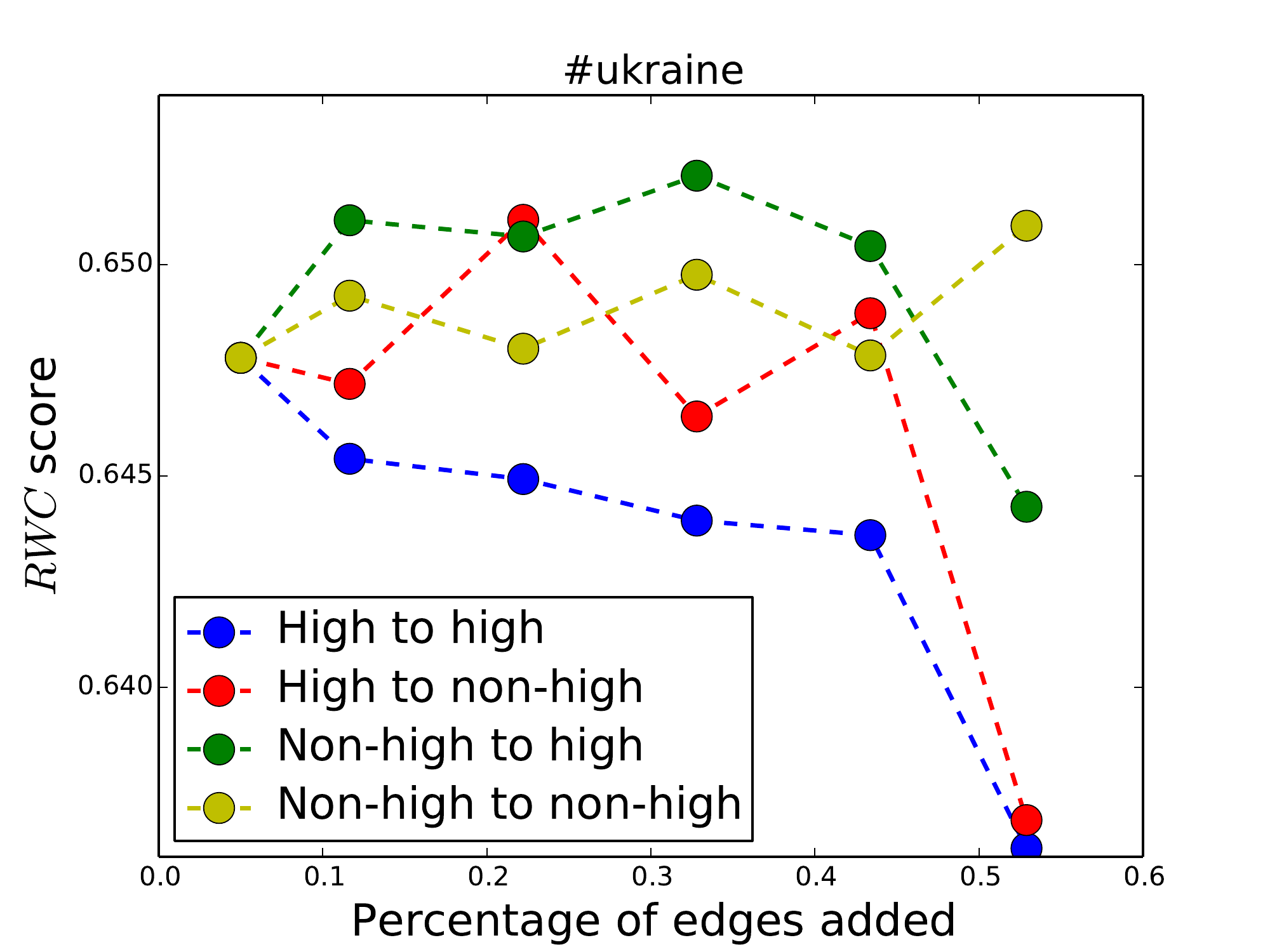}}
\end{minipage}\par\medskip
\caption{Comparison of different edge-addition strategies after the addition of 50 edges.}
\label{fig:edge_addition_high_nonhigh}
\vspace{-\baselineskip}
\end{figure*}

In this section, we evaluate different edge-addition strategies. 
The goal is to test the hypothesis that adding edges among high-degree nodes 
on the two sides of the controversy gives the highest decrease in polarity score. 
For each of the 10 datasets, we generate a list of random high-degree nodes and non high-degree nodes on each side. 
We then generate a list of 50 edges, drawn at random from the sampled nodes, 
and corresponding to the 4 possible combinations 
(high/non high to high/non high edges). 
The results of these simulations are shown in Figure~\ref{fig:edge_addition_high_nonhigh}. 
We see that, 
despite the fact that high-degree nodes are selected at random, 
connecting such nodes gives the highest decrease in polarity score (blue line).

\subsection{Case study}
\label{sec:casestudy}
In order to provide qualitative evidence on the functioning of our algorithms 
on real-world datasets, 
we conduct a case study on three datasets.
The datasets are chosen for the ease of the interpretation of the results,
since they represent topics of wider interest 
(compared to {\it beefban}, for example, which is specific to India). 

The results of the case study are summarized in Table~\ref{tab:casestudy}.
We can verify that the recommendations we obtain are meaningful
and agree with our intuition for the proposed methods.
The most important observation is that 
when comparing {\rov} and {\rovap}
we see a clear difference in the type of edges recommended. 
For example, for {\it obamacare}, {\rov} recommends edges 
from {\it mittromney} to {\it barackbobama}, 
from {\it barackobama} to {\it paulryanvp} (2012 vice president nominee), etc. 
Even though these edges make sense in connecting opposing sides, 
they might be hard to materialize in the real world. 
This issue is mitigated by {\rovap}, 
which recommends edges between less popular users, 
yet still connecting opposing viewpoints.
Examples include the edge ({\it csgv}, {\it dloesch}) for guncontrol, 
which connects a pro-gun-control organization to a conservative radio host, 
or the edge ({\it farhankvirk}, {\it pamelageller}), 
which connects an islamist blogger with a user who wants to 
``Stop the Islamization of America.''\footnote{Note that since some of the data is from 2012-13, 
some accounts may have been deleted/moved 
(e.g., {\it paulryanvp}, {\it truthteam2012}, etc). 
Also, some accounts may have changed stance in these years. 
Interested readers can use the internet archive Wayback machine to have a look at past profiles.}

Additionally, we provide a quantitative comparison of the output 
of the two algorithms, {\rov} and {\rovap},
by extracting various statistics regarding the recommended edges
In particular we consider:
($i$) {\em Total number of followers}.
We compute the median number of followers from all edges suggested by {\rov} and {\rovap}. 
A high value indicates that the users are more central.
($ii$) {\em Overlap of tweet content},
For each edge we compute the Jaccard similarity of the text of the tweets of the two users. 
We aggregate these values for each dataset, 
by taking the median among all edges.
A higher value indicates that there is higher similarity between the tweet texts 
of the two users recommended by the algorithm.
($iii$) {\em Fraction of common retweets}.
For each recommended edge $(x,y)$, 
we obtain all other users who retweeted users $x$ and $y$, and 
compute the Jaccard similarity of the two sets. 
As before,
we aggregate for each dataset, by taking the median among all edges. 
A higher value indicates that there is a higher agreement in endorsement for users $x,y$ on the topic.

The results are presented in Table~\ref{tab:casestudy2}. 
We observe that the results agree with our intuition. 
For example, {\rovap} produces edges with a lower number of followers (not extremely popular users), 
who have more common retweets and a higher overlap in terms of tweet content.

To provide a visual understanding on how our algorithms help reduce controversy, we visualized the two sides of a controversial topic before and after adding edges as proposed by our algorithm {\rovap}. We used Gephi's force-directed layout algorithm for a similar number of iterations on each network. Figure~\ref{fig:visualization} shows the results on two datasets \#beefban and \#russia\_march, before (\ref{fig:visualization}(a,c)) and after (\ref{fig:visualization}(b,d)) the addition of 20 edges suggested by {\rovap}. The end points of the new edges are marked by green and black colors. 
We can clearly see that after the addition of the edges by our algorithm, the two sides appear to come closer.

\begin{table*}[t]
\caption{\label{tab:casestudy}
Twitter handles of the top edges picked by our algorithms for different datasets. 
}
\centering
\small
\begin{tabular}{l l l l l l l}
\toprule
 & \multicolumn{2}{c}{obamacare} & \multicolumn{2}{c}{guncontrol} & \multicolumn{2}{c}{\#netanyahuspeech} \\
\cmidrule(lr){2-3} \cmidrule(lr){4-5} \cmidrule(lr){6-7}
& node1 & node2 & node1 & node2 & node1 & node2 \\
\midrule
\multirow{5}{*}{{\rov}} & mittromney & barackobama & ghostpanther & barackobama & maxblumenthal & netanyahu \\
 & realdonaldtrump & truthteam2012 & mmflint & robdelaney & bipartisanism & lindasuhler \\
 & barackobama & drudge\_report & miafarrow & chuckwoolery & harryslaststand & rednationrising \\
 & barackobama & paulryanvp & realalexjones & barackobama & lindasuhler & marwanbishara \\
 & michelebachmann & barackobama & goldiehawn & jedediahbila & thebaxterbean & worldnetdaily \\
\midrule
\multirow{5}{*}{{\rovap}} & kksheld & ezraklein & chuckwoolery & csgv & farhankvirk & pamelageller \\
 & lolgop & romneyresponse & liamkfisher & miafarrow & medeabenjamin & annebayefsky \\
 & irritatedwoman & motherjones & csgv & dloesch & 2afight & sttbs73 \\
 & hcan & romneyresponse & jonlovett & spreadbutter & rednationrising & palsjustice \\
 & klsouth & dennisdmz & drmartyfox & huffpostpol & jvplive & chucknellis \\
\bottomrule
\end{tabular}
\vspace{-\baselineskip}
\end{table*}

\begin{table}
\caption{\label{tab:casestudy2}
Quantitative comparison of recommendations
from {\rov} and {\rovap}. 
$^*$ indicates that the result is statistically significant with $p<0.1$, 
and $^{**}$ with $p<0.001$. 
Significance is tested using Welch's $t$-test for inequality of means.
}
\centering
\begin{tabular}{c c c}
\toprule
& {\rov} & {\rovap} \\
\cmidrule(lr){2-3}
NumFollowers & 50729 & 36160$^*$\\
ContentOverlap & 0.054 & 0.073$^{**}$ \\
CommonRetweets & 0.029 & 0.063$^{**}$ \\
\bottomrule
\end{tabular}
\vspace{-\baselineskip}
\end{table}

\subsection{Time performance}
\label{sec:timetaken}

\begin{figure}
\centering
\includegraphics[width=0.40\textwidth, clip=true, trim=0 0 10 35]{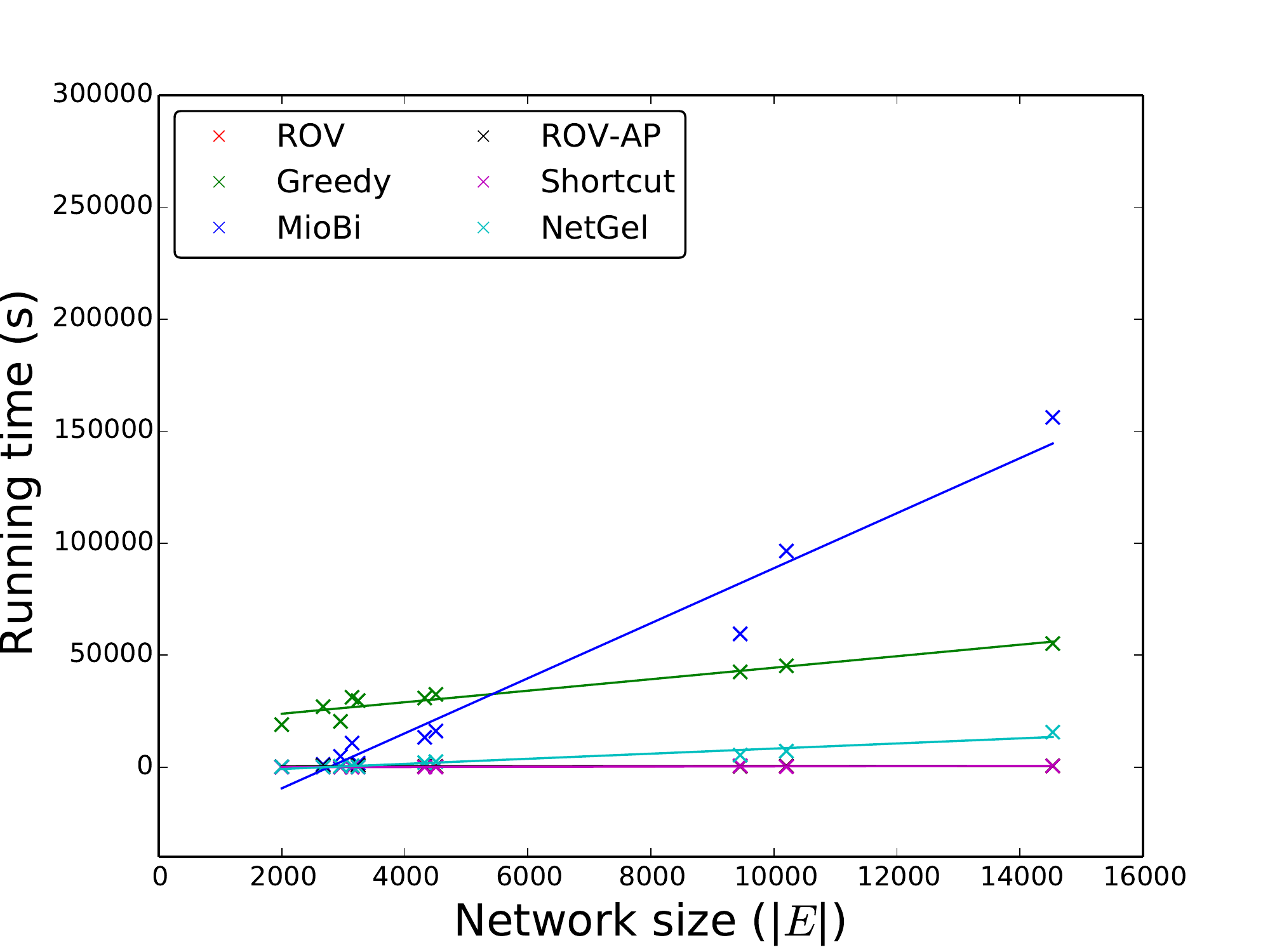}
\caption{Running time of the proposed algorithms and competitors.
{\rov} and {\rovap} almost overlap.}
\label{fig:timetaken}
\end{figure}

Finally, we measure the performance of our algorithms in terms of running time. 
In Figure~\ref{fig:timetaken}, we see that both our algorithms {\rov} and {\rovap} are fast 
in comparison with other related approaches. Greedy and MioBi are the slowest.

Moreover, we compute the improvement in running time 
due to the incremental computation of Section~\ref{sec:recomputation}.
The speedup compared to the non-incremental version of the algorithm, 
for the different datasets, ranges from 5 to 65x.
In general, the speedup is larger for denser graphs.

\section{Conclusions}

We considered the problem of bridging opposing views on social media by recommending relevant content 
to certain users (edges in the endorsement graph).
Our work builds on recent studies of controversy 
in social media and use a random walk-based score as a measure of controversy.
We first proposed a simple, yet efficient, algorithm to bridge opposing sides.
Furthermore, 
inspired by recent user studies on how users prefer to consume content from opposing views, 
we improved the algorithm to take into account the probability of an edge being accepted. 
Finally, we also proposed a way to incrementally compute the random-walk score 
using matrix operations, which typically gives more than an order of magnitude improvement in runtime.
We evaluated our algorithms on a wide range of real-world datasets in Twitter, 
and showed that our methods outperform other baselines.

%


\textbf{Future work.}
Our approach relies on the random-walk based optimization function~\cite{garimella2016quantifying}.
Although this measure has been proven to be effective it has a few drawbacks. 
In particular, the measure is applicable to controversies having two sides.
One way to overcome this restriction is to assume the presence of multiple clusters, and 
define the measure accordingly.
In the future, we plan to experiment with this generalization of our method, 
as well as, investigate the edge-recommendation problem for other objective functions. 

As mentioned in Section~\ref{section:related-work}, 
previous work deals mostly with the problem of {\em how} to connect opposing sides, 
while our work provides methods for selecting {\em who} to recommend.
Another interesting direction is studying the problem of {\em what} to recommend.

\mpara{Acknowledgements}
This work is supported by the European Community's H2020 Program under the
scheme ``INFRAIA-1-2014-2015: Research Infrastructures,'' grant agreement
\#654024 ``SoBigData: Social Mining \& Big Data Ecosystem.''


\bibliographystyle{abbrv}
\bibliography{biblio}
\balancecolumns 

\end{document}